\documentclass[useAMS,usenatbib,a4paper]{mn2e}

\usepackage{footmisc}
\usepackage{hyperref}

\usepackage{amsmath}
\usepackage{amssymb}
\usepackage{amstext}

\usepackage{mathtools}
\usepackage{aas_macros}
\usepackage{graphicx}
\graphicspath{ {./images/} }
\usepackage{natbib}
\title[Pressure Broadening in Exoplanetary Atmospheres]{Effect of Pressure Broadening on Molecular Absorption Cross Sections in Exoplanetary Atmospheres}
\author[Hedges and Madhusudhan]{Christina Hedges\thanks{E-mail: chedges@ast.cam.ac.uk} \& 
  Nikku Madhusudhan\thanks{E-mail: nmadhu@ast.cam.ac.uk}
  \\
  Institute of Astronomy, University of Cambridge, Madingley Road, Cambridge, CB3 0HA, UK \\
}

\begin{document}

\date{Accepted 2016 January 29. Received 2016 January 29; in original form 2015 December 3}

\label{firstpage}
\pagerange{\pageref{firstpage}--\pageref{lastpage}}
\maketitle

\begin{abstract}
Spectroscopic observations of exoplanets are leading to unprecedented constraints on their atmospheric compositions. However, molecular abundances derived from spectra are degenerate with the absorption cross sections which form critical input data in atmospheric models. Therefore, it is important to quantify the uncertainties in molecular cross sections to reliably estimate the uncertainties in derived molecular abundances. However, converting line lists into cross sections via line broadening involves a series of prescriptions for which the uncertainties are not well understood. We investigate and quantify the effects of various factors involved in line broadening in exoplanetary atmospheres - the profile evaluation width, pressure versus thermal broadening, broadening agent, spectral resolution, and completeness of broadening parameters - on molecular absorption cross sections. We use H$_2$O as a case study as it has the most complete absorption line data. For low resolution spectra (R$\lesssim$100) for representative temperatures and pressures (T $\sim$ 500K-3000K, P$\lesssim$1 atm) of H$_2$-rich exoplanetary atmospheres we find the median difference in cross sections ($\delta$) introduced by various aspects of pressure broadening to be $\lesssim$1\%. For medium resolutions (R$\lesssim$5000), including those attainable with JWST, we find that $\delta$ can be up to 40\%. For high resolutions (R$\sim$10$^5$) $\delta$ can be $\gtrsim$100\%, reaching $\gtrsim$1000\% for low temperatures (T$\lesssim$500K) and high pressures (P$\gtrsim$1 atm). The effect is higher still for self broadening. We generate a homogeneous database of absorption cross sections of molecules of relevance to exoplanetary atmospheres for which high temperature line lists are available, particularly H$_2$O, CO, CH$_4$, CO$_2$, HCN, and NH$_3$. 

\end{abstract}
\section{Introduction}

In recent years it has become possible to observe high-precision atmospheric spectra of a variety of exoplanets detected via transits, direct imaging and Doppler surveys \citep[see e.g. review by][]{madhusudhan2014}. For example, high-precision observations with the HST Wide Field Camera 3 (WFC3) in the near-infrared (1.1-1.7 $\mu$m) has led to unambiguous detections of H$_2$O in several hot Jupiter atmospheres \citep{2013ApJ...774...95D,mcCullough2014,2014ApJ...793L..27K,2014ApJ...791L...9M}. Despite the modest resolution (R$\sim$10-100; depending on spectral binning), the high photometric precisions of HST WFC3 spectra have allowed unprecedented constraints on the H$_2$O abundances in these atmospheres \citep[e.g.][]{2014ApJ...793L..27K,2014ApJ...791L...9M}. On the other hand, it has also become possible to detect molecules such as H$_2$O and CO in atmospheres of hot Jupiters orbiting bright stars using very high resolution ($R = 10^5$) infrared Doppler spectroscopy with large ground-based facilities \citep{snellen2010,brogi2012}. Complementary to short period exoplanets, high-resolution and high-precision spectra have also been reported for young giant planets on large orbital separations discovered via direct imaging from ground leading to detections of several key molecules including H$_2$O, CO and CH$_4$ \citep{konopacky2013}. While these observational advancements are already leading to detailed constraints on the chemical compositions and physical processes in exoplanetary atmospheres, the field will be further revolutionised with upcoming facilities including JWST ($R\sim1000-3000$) in space and several large ground-based facilities such as the E-ELT ($R \sim 10^5$). 

Central to the interpretation of such observations, however, is the accuracy of the fundamental inputs to atmospheric models. The opacity contributing to a spectrum due to a given chemical species is proportional to the product of its absorption cross section and the molar abundance of that species. Therefore, the chemical abundances derived from atmospheric spectra, using standard retrieval codes \citep[e.g.][]{madhusudhan2009,madhusudhan2011,line2013a,benneke2012,lee2012,waldmann2015}, are directly degenerate with the absorption cross sections. In conventional atmospheric models the absorption cross sections are fixed as fundamental inputs and hence the uncertainties on derived chemical abundances assume no uncertainties in cross sections. However, with improving data quality, both in precision and resolution, it becomes imperative to examine the uncertainties in these model inputs. Ultimately the precision of a derived molecular abundance will be limited by the accuracy of its wavelength-dependent cross sections as they are assumed as model inputs.

Significant progress has been made in recent years to generate molecular absorption line lists for molecules of relevance to exoplanetary atmospheres. While traditionally molecular line databases such as HITRAN (Rothman et al. 2006) provided data for a large compendium of molecules, the data was typically available for terrestrial applications of temperatures $\lesssim$ 300 K. Since exoplanetary atmospheres that are observable with current instruments span much higher temperatures ($\sim$600 - 3000 K) it has become necessary to generate high temperature line lists for many molecules. Therefore, several new high-temperature molecular line lists have been reported in recent years. For example, the recently revised HITEMP database (Rothman et al. 2010) provides a compilation of theoretical and experimental high temperature line lists and line broadening parameters for important molecules such as H$_2$O, CO, and CO$_2$. More recently, and more extensively, the ExoMol database (Tennyson \& Yurchenko 2012) has reported high temperature theoretical line lists for numerous molecules of relevance to exoplanetary atmospheres, including the largest line lists for CH$_4$ and NH$_3$ known to date (e.g. Yurchenko et al. 2013). 

Molecular absorption cross sections are generated from transition line lists by incorporating the appropriate line broadening and binning to the required resolution. Several factors can cause line broadening in exoplanetary atmospheres, such as thermal Doppler broadening and pressure broadening. However, while detailed high temperature line lists are becoming available for several molecules, there is still a lack of detailed line broadening parameters that are required for generating accurate cross sections. Freedman et al. (2008) and Freedman et al. (2014) report opacity calculations using existing high-temperature line lists and highlight the lack of detailed pressure broadening parameters. In particular, current atmospheric observations are most sensitive for giant exoplanets with H$_2$-rich atmospheres, and hence molecular opacities in models need to incorporate pressure broadening due to H$_2$. However, while latest databases such as HITEMP provide broadening data for self and air broadening, H$_2$ broadening data is currently unavailable for most molecules. Thus, high-temperature molecular cross sections for several molecules rely on insufficient H$_2$ pressure broadening data (Freedman et al. 2008, 2014; Sharp \& Burrows 2008). On the other hand, purely theoretical line lists such as ExoMol are computed in zero pressure conditions due to which pressure broadening data is typically unavailable and hence not included while computing cross sections \citep[e.g.][]{2013Icar..226.1673H}. Finally, even where broadening parameters are available, deriving cross sections from line lists involves a series of numerical considerations which can influence the cross sections. 

In the present work, we investigate the dependence of molecular cross sections on the various factors involved in computing them from molecular line lists including both thermal and pressure broadening. We discuss each aspect of the construction of the cross sections and where errors can be introduced by comparing different data sources, temperatures, pressures, broadening agents, evaluation widths and resolutions. Meaningful comparisons can be difficult to make as cross sections span many orders of magnitude and contain many peaks and troughs in features across wavelengths. As discussed in section~\ref{measuring} we find the median percentage difference to be a useful measure and representative of the change due to pressure broadening as a whole. To assess this we also generate a bank of cross sections with only thermal broadening included for comparison with those including pressure broadening. Using this method, we create a database of molecular cross sections from a range of available line list sources. The cross sections generated span a wide range of pressures (10$^{-4}$ - 100 atm) and temperatures (300 - 3500 K) that are relevant for exoplanetary atmospheres and sub-stellar objects. 

We investigate the cross sections over a wide range in spectral resolution (R=10$^2$-10$^5$) which reflects current and future instruments for atmospheric characterisation of exoplanets as discussed above. There are a wide array of future instruments that we can prepare for. The James Webb Space Telescope (JWST) is scheduled for launch in 2018 and will host several instruments in the infra-red range. NIRSPEC and MIRI spectrographs will hold significance for the characterisation of exoplanet atmospheres. NIRSPEC will encompass the 0.6-5 $\mu$m range using three overlapping bands. MIRI will be a particularly broadband instrument stretching from 5-20 $\mu$m in wavelength. NIRSPEC will have a spectral resolution ranging up to R=1400-3600 $\mu$m for its highest resolution grating and MIRI will achieve R $\approx$ 3000 $\mu$m. This is a drastic increase in our current capabilities with HST which warrants an investigation into our current modelling inputs. In the distant future we also anticipate the E-ELT to have R=100,000 and be highly capable of atmospheric characterisation for exoplanets. Based on these instruments we present our tests of cross sections at a range of resolutions of R=100, 1000, 3000, 10,000 and 100,000. This gives a representative picture of what we can achieve now and in the future. 

\section{Line List Sources}
\label{LineListDatabases}

There are several line list repositories available where lists of transitions for molecules can be obtained. Those most relevant to atmospheric characterisation and that are publicly available are given in Appendix \ref{LineListSources} including the number of transitions each list contains. These databases are given in a variety of formats but each source contains the key parameters (e.g. Einstein coefficients, degeneracies, energy levels, etc.) and a method for obtaining the line positions (e.g. in wavenumber, $\nu$) and the line intensities $S(T)$ for a given temperature (T). These line data sources cover a wide spectral range, typically spanning the visible to mid-IR (e.g. $\sim$0.5 $\mu$m to $\sim$30 $\mu$m), although this can vary between the different line lists. Line transitions are uniformly spread in frequency space leading to line lists being given in wavenumber rather than wavelength. Sources also range in their completeness with some containing fewer transitions than others. Lack of completeness leads to less reliable cross section data for two reasons; gaps in the wavenumber coverage cause some features to be missed from the cross section and lines of lower intensity which can contribute significantly to the cross section are not represented. As such accurate cross sections require the most complete lists of molecular transitions.

One of the largest and most well established of all the repositories for molecular line lists is the HITRAN database, which has been updated every few years \citep{1998JQSRT..60..665R,2013JQSRT.130....4R}. The HITRAN database has been mainly used for terrestrial applications and predominantly includes molecules of importance for the Earth's atmosphere, with temperatures below $\sim$300 K.  Because of this lower temperature approach these data are less applicable for the most observable exoplanetary atmospheres. For current atmospheric observations of exoplanets we require spectroscopic data covering a higher temperature range, e.g. for applications to highly irradiated planets and young directly-imaged planets which have a wide range of temperatures going up to 3000 K. This need has been met by the newer HITEMP database, also known as HITEMP2010 to distinguish from an earlier version, which contains fewer molecules but many more transitions for each \citep{2010JQSRT.111.2139R}. HITEMP currently covers OH, NO, CO, CO$_2$ and H$_2$O which are particularly useful for hot Jupiter atmospheres, accurate up to temperatures of 4000K. More recently, the ExoMol database has begun addressing the deficit of data available for molecules of astrophysical importance at high temperatures \citep{2012MNRAS.425...21T}.

\section{Line Broadening}
\label{sec:broadening}
The diversity of physical conditions in exoplanetary atmospheres can lead to different types and degree of line broadening. Exoplanetary atmospheres span a wide range of temperatures ($\sim$400 - 3000 K) and dynamical parameters (e.g. wind speeds, and orbital and spin rotation rates) ranging from tidally locked close-in planets to young giant planets on wide orbital separations. The two prominent sources of line broadening in planetary atmospheres are thermal (Doppler) broadening and pressure (collisional) broadening \citep{1978tpai.book.....C,1978ApJ...220.1001M,2010ARA&A..48..631S}. Thermal Doppler broadening is caused by the line-of-sight thermal velocity distribution of molecules at a given temperature in the planetary atmosphere. Pressure broadening is induced by collisions between chemical species with the collision frequency being a strong function of pressure. 

Other sources of broadening can be prevalent depending on the planetary properties and observing geometry. In principle, natural broadening due to the intrinsic uncertainty in energy levels is always present but is expected to be negligible compared to other broadening mechanisms discussed above. Further broadening and shifting of spectral lines can be caused by the planetary rotation and strong winds in the planetary atmosphere, especially for close-in hot Jupiters observed in transmission spectra \citep{2007ApJ...669.1324S,2012ApJ...751..117M,2013cctp.book..277S}. Finally, rotational broadening due to the spin of the planet can also be significant, especially for exoplanets that are not tidally locked such as those on wide orbital separations \citep{2014Natur.509...63S}. These sources of broadening can be important on a case-by-case basis. 

\subsection{Broadening Profiles}
\label{numericalmethod}

Opacities used in models of exoplanetary atmospheres typically include thermal and pressure broadening of the spectral lines wherever the broadening parameters are available. While thermal broadening is straightforward to include (e.g. \cite{2013Icar..226.1673H}), the parameters for pressure broadening are sparse as relevant approximations need to be made for each molecule considered \citep{Freedman,2008ApJS..174..504F}, as discussed in section~\ref{sec:availability}. In the present work, we consider both thermal and pressure broadening and investigate the effect of the assumed pressure broadening parameters on the resulting absorption cross sections. 

Under the assumption of a Maxwell-Boltzmann thermal velocity distribution the Doppler broadening takes the form of a Gaussian profile. On the other hand, pressure broadening is represented by a Lorentzian profile. The Doppler and Lorentzian broadening profiles are given, in wavenumbers in cm$^{-1}$, as 
\begin{equation} f_D(\nu-\nu_0)=\frac{1}{\gamma_G\sqrt{\pi}}exp(-\frac{(\nu-\nu_0)^2}{\gamma_G^2})\end{equation}
\begin{equation} f_P(\nu-\nu_0)=\frac{1}{\pi}\frac{\gamma_L}{(\nu-\nu_0)^2+\gamma_L^2}\end{equation}

where, $\nu_o$ is the centroid in wavenumbers, $\gamma_G$ we define as the Doppler width and $\gamma_L$ is the Lorentzian pressure broadening half-width at half-maximum (HWHM) both in units of cm$^{-1}$. These are given by \citep[see e.g.][]{2013Icar..226.1673H,1998JQSRT..60..665R}:  

\begin{equation}\label{eq:gausswidth} \gamma_G=\sqrt{\frac{2k_BT}{m}}\frac{\nu_0}{c}\end{equation}
\begin{equation}\label{eq:lorwidth} \gamma_L=\Big(\frac{T_{ref}}{T}\Big)^n\;P\;\sum_b\gamma_{L,b}\;p_{b} \end{equation}

where, $P$ is pressure in atm, $T$ is the temperature in Kelvin, $T_{ref}$ is the reference temperature (usually 296K), $p_b$ is the partial pressure of the broadener, $n$ is a temperature scaling factor and $\gamma_{L,b}$ indicates the Lorentzian HWHM due to a specific broadening molecule in units of cm$^{-1}$/atm. Here $k_B$ is the Boltzmann constant, $m$ is the mass of the molecule in grams and $c$ is the speed of light in cm/s. Here, $\sum$ signifies the sum over all the broadening parameters for each broadening medium. 

Pressure broadening is typically harder to evaluate than thermal broadening for multiple reasons. Firstly, as alluded to above, the line-by-line pressure broadening parameters, $n(\nu)$ and $\gamma(\nu)$, are typically unavailable for most molecules under the conditions encountered in exoplanetary atmospheres, e.g. high temperatures up to $\sim$3000 K and varied atmospheric compositions such as H$_2$-rich gas giant atmospheres (Freedman et al. 2008,2014). We discuss this further in section \ref{sec:availability}. Secondly, the Lorentzian profile contributes a higher percentage to its extensive wings which can result in a significant amount of the intensity being moved into the wings of the line profile. In cases of extreme broadening this can significantly increase the impact of high intensity transitions far from the line centre and influence cross sections from neighbouring low intensity transitions. Therefore, the profile needs to be treated particularly carefully by sampling the Voigt well to approximate the profile shape with low uncertainties. This ensures appropriate normalisation so that no intensity is lost from binning the profile. 

\begin{figure}
  \centering
  \includegraphics[scale=0.3]{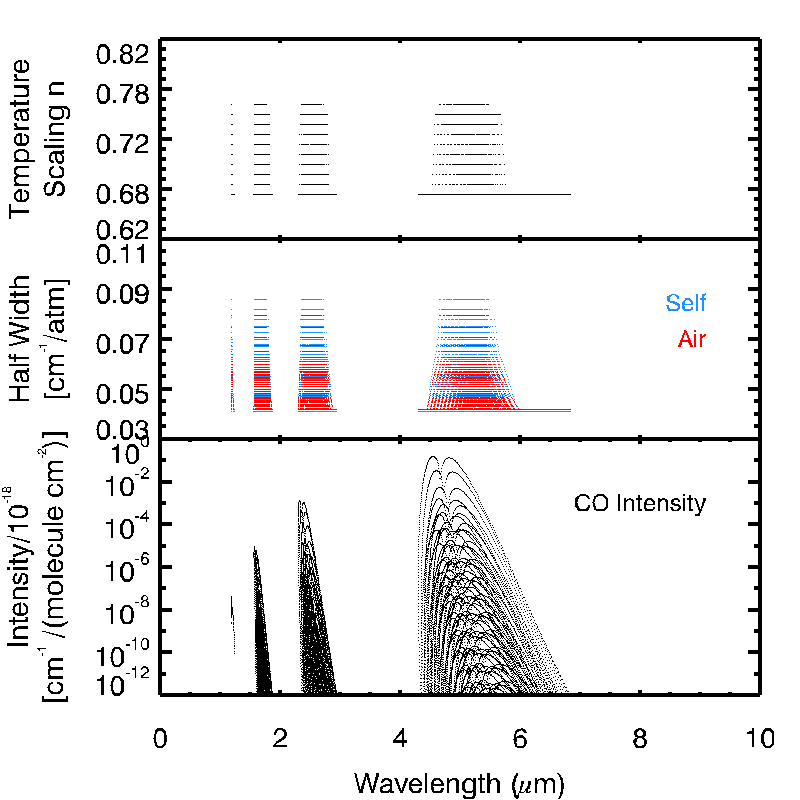}
  \caption{Pressure broadening parameters for the CO line list from HITEMP. We see discrete values of each parameter as the CRB formalism uses the discrete J quantum number. Self and air broadening parameters have both been plotted here and show that in general air broadening values are slightly lower than self broadening.}
  \label{ComparePresParams}
\end{figure}

Figure \ref{ComparePresParams} shows the spread of both the Lorentzian HWHM ($\gamma_{L,b}$) and the temperature scaling parameter (n) with the intensity for CO, one of the simplest molecules in the database. We can see discrete levels in each parameter due to the discrete nature of the J quantum number that the pressure broadening values are generated from in Complex Robert-Bonamy (CRB) calculations \cite[]{Antony2006,Robert1979}. The spread is quite narrow in this parameter space leading us to believe that a mean value of each parameter could reasonably approximate the results of a detailed line-by-line treatment of pressure broadening.

\begin{figure}
  \centering
  \label{ptregions}
  \includegraphics[scale=0.45]{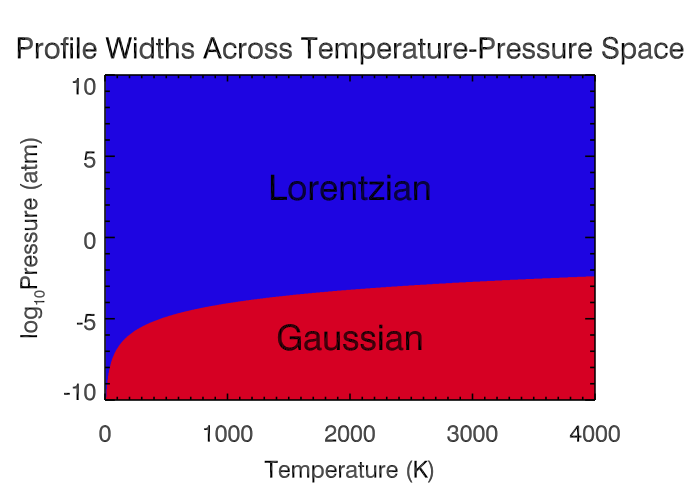}
  \caption{Comparison of widths of line cores of Gaussian vs Lorentzian profiles in pressure-temperature space. The red (blue) region represents P-T space where the HWHM of a Gaussian (Lorentzian) profile is wider than that of a Lorentzian (Gaussian) profile.  The Gaussian profile is wider at low pressures and the Lorentzian at high pressures, as expected.} 
  \label{VoigtSurface}
\end{figure}

It is useful to understand where each of the two broadening mechanisms contributes most significantly in the pressure-temperature ($P$-$T$) space relevant for exoplanetary atmospheres. However, as the two profiles are disparate with a Gaussian profile containing more information in the core compared to the extended Lorentzian wings it is difficult to compare the two with a common metric. Nevertheless, Figure~\ref{VoigtSurface} shows a comparison between the Doppler HWHM ($\gamma_G\sqrt{ln2}$) and the Lorentzian HWHM over the $P$-$T$ space of interest to give an approximation of where each profile contributes most significantly. As expected, at low pressures thermal (Gaussian) broadening provides a significant contribution to the final profile core, whereas at high pressures pressure (Lorentzian) broadening is stronger. Closer to the boundary between these two regimes, both broadening mechanisms are likely to contribute significantly to core of the profile. Furthermore, due to the extended wings of the Lorentzian it is generally advisable to consider both broadening contributions even when the Lorentzian HWHM is narrow in comparison to the Gaussian. This is done using the Voigt profile discussed below. 

As discussed in \citet{2012RSPTA.370.2495N} it has been shown through comparisons with experimental data on pressure broadening that there is some deviation in reality from the Voigt profile. This is due to the change in velocity of the broadener particle by the collision with the broadening agent which affects the profile shape, the width and the shift of the line. Here we have used only the standard Voigt profile and have not investigated further in terms of profile shape. It would be possible to change to a more sophisticated profile and regenerate molecular cross sections if it were found to be an important factor. These deviations due to the velocity of the particles undergoing collisions are of the order of a few percent and we do not expect a more physically accurate Voigt profile shape to impact our results.

\subsection{Evaluating the Voigt Profile}
\label{numericalmethod}

The joint contributions due to thermal and pressure broadening are modelled using a Voigt profile which is a  convolution of the Gaussian and Lorentzian profiles, given as 

\begin{equation} f_V (\nu-\nu_0) = \int_{-\infty}^{\infty}f_G(\nu^\prime-\nu_0)f_L(\nu-\nu^\prime)d\nu^\prime .
\end{equation}

The characteristic width of the Voigt function is investigated by \cite{Olivero1977233}. Using coefficients from this work we define the Voigt width $\gamma_V$ with the approximation

\begin{equation}\label{eq:voigtwidth} \gamma_V\approx 0.5346 \gamma_L+\sqrt{0.2166\gamma_L^2+\gamma_G^2}.\end{equation}

This Voigt width is used in later sections to approximate the width of the combination of the two profile types.

The issue of how best to evaluate the Voigt profile is a well known problem. The profile must be calculated accurately and quickly for a wide range of Lorentzian and Gaussian profiles, corresponding to the wide range of temperature and pressure values, for potentially millions of lines of a given molecule. The two parameters used for generating the profile are 
\begin{equation} u=\frac{\nu-\nu_0}{\gamma_G},~~~ 
a=\frac{\gamma_L}{\gamma_G}\end{equation}
where, $u$ is the distance from the profile centroid and $a$ is the ratio of the Lorentzian and Gaussian widths \citep{2007MNRAS.375.1043Z}. To be able to calculate the Voigt function accurately $u$ must be evaluated over many orders of magnitude to encompass the relevant temperature and pressure region.

The Voigt function is given as
\begin{equation}
f_v(\nu,\gamma_G,\gamma_L)=H(a,u)
\end{equation}

where, as above $\gamma_G$ and $\gamma_L$ are the widths of the Gaussian and Lorentzian widths and $\nu$ denotes wavenumber. Here, 
\begin{equation}
H(a,u)=\frac{1}{\pi}\int_{-\infty}^\infty\frac{ae^{-t^2}}{(u-t)^2+a^2}dt.
\end{equation}

From formula 7.4.13 in \cite{abramowitz1964handbook}
\begin{equation}
\int_{-\infty}^\infty \frac{ye^{-t^2}}{(x-t)^2+y^2}dt=\pi \mathbb{R}w(x+iy), 
\end{equation}

where 

\begin{equation}
w(z)=e^{-z^2}erfc(-iz), 
\end{equation}

where erfc is the complimentary error function

\begin{equation}
erfc(z)=1-erf(z), 
\end{equation}

where erf is the error function

\begin{equation}
erf(z)=\frac{2}{\sqrt{\pi}}\int_0^ze^{-t^2}dt.
\end{equation}

From this we can see that 
\begin{equation}
f_v(\nu,\gamma_G,\gamma_L)=\mathbb{R}w(u+ia)
\end{equation}

The function $w(z)$ is known as the Faddeeva function and here is calculated using the Faddeeva package \citep{Faddeeva}.

Several numerical methods have been proposed to compute the Voigt profile accurately and efficiently over different regions of parameter space \citep{Schreier20111010}. We implement the Voigt profile using a method based on the complex error function or Faddeeva function \citep{Faddeeva}. This is a fast and accurate method with the relevant libraries publicly available. The Faddeeva package includes \emph{Algorithm 916}\ from \cite{DBLP:journals/corr/abs-1106-0151} which is known to provide accurate results. We find this package gives converged profiles over our required parameter space with fast computational speeds of less than 2 ms per profile. We use a sampling rate of 6 points per Voigt width, which is much finer than our required final resolution. We evaluate the Voigt profile to 500 Voigt widths around the centroid to accurately capture the information in the Lorentzian wings, as the derived cross sections are critically dependent on this evaluation width (discussed in more detail in section ~\ref{EvaluationWidth}). 

\subsection{Availability of Broadening Parameters}
\label{sec:availability}

\begin{figure}
  \centering
  \includegraphics[scale=0.5]{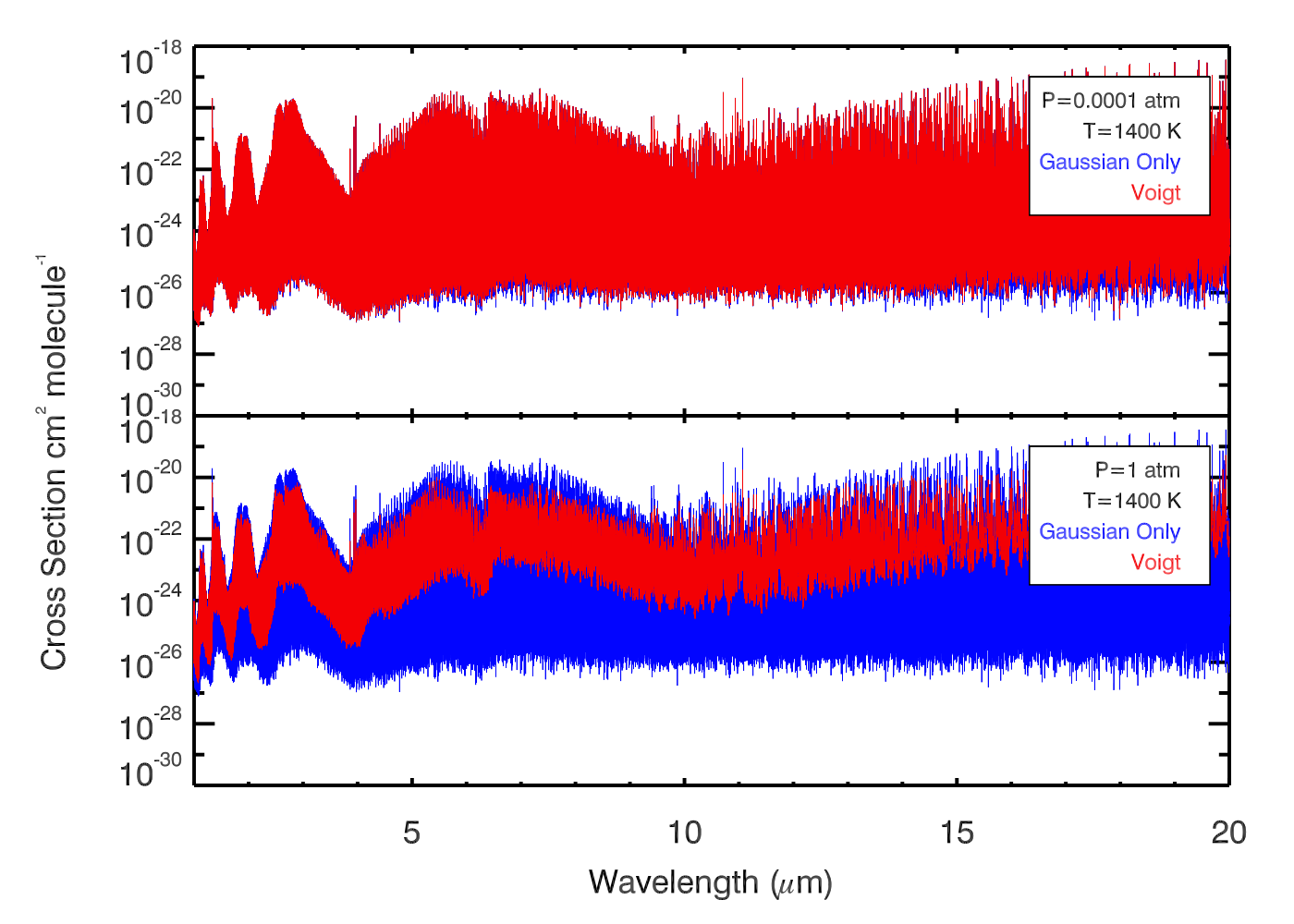}
  \caption{Comparison of purely thermally-broadened H$_2$O cross sections (blue) with cross sections including both thermal and pressure broadening using a Voigt profile (red) at the native line spacing of 0.01 cm$^{-1}$. A combination of the two broadening types brings extensive wings from the Lorentzian component which brings up the level of the continuum.}
  \label{fig:voigtvsgauss}
\end{figure}

Computing line broadening, as discussed above, requires broadening parameters for each line for both the thermal and pressure broadening components. For thermal broadening of a spectral line of a given molecule at a given temperature the Doppler width ($\gamma_G$) is easily calculated from Equation \ref{eq:gausswidth}. Several recent works have reported such thermal broadened cross sections for several molecules, e.g. \cite{2006MNRAS.368.1087B}, \cite{2013Icar..226.1673H}. Figure~\ref{fig:voigtvsgauss} shows a comparison between cross sections for H$_2$O generated with only thermal broadening and those generated with a full Voigt profile including both thermal and pressure broadening. It is clear that the Voigt profile has a significant effect on the low intensity lines, and increases the overall continuum of the molecular cross sections, especially where pressure is high.

Despite their critical importance pressure broadening parameters are not yet readily available for all molecules of relevance to exoplanetary atmospheres (Freedman et al. 2008,2014). As shown in Equation \ref{eq:lorwidth}, the parameters required for computing pressure broadening are the Lorentzian HWHM ($\gamma_L$) for the required broadeners and the temperature-scaling parameter ($n$) for each line in the line list. These parameters are hard to determine. 
Theoretical calculations of pressure broadening parameters are particularly time consuming and not covered for a wide range of molecules or broadening agents, particularly at high temperatures. Such methods have been explored by \cite{1997JQSRT..57..485G,2011Icar..213..720G} and used to generate the HITRAN database, e.g. Complex Robert-Bonamy (CRB) calculations are used as discussed in \cite{Gamache1998319,Gamache2012976} where values are also verified experimentally. Molecular line lists in the HITRAN data base do contain pressure broadening parameters for self-broadening and air-broadening, but are typically relevant only to low temperature atmospheres ($\sim$300 K). For high temperature exoplanetary atmospheres, particularly of H$_2$-rich atmospheres that are most observable, pressure broadening data is still scarce. Typically, state-of-the-art ab initio line lists such as ExoMol are computed under zero pressure conditions due to which the pressure broadening line parameters are not available. On the other hand experimental data are also scarce for the conditions relevant for exoplanetary atmospheres.

In recent years, significant efforts have been dedicated towards generating pressure broadening parameters for exoplanetary applications, particularly with a focus on broadening molecules such as $H_2$ and other molecules of interest outside of Earth applications. Useful parameters for important molecules can be found in works such as \citep{2015ApJS..216...15L} and \citep{Faure201379} where pressure broadening coefficients are made available. In this work we use only air and self broadening provided by HITEMP and HITRAN and the PS 1997 list for water from \citep{PS}, which contains H$_2$ broadening R. Freedman, personal communication). We focus on the H$_2$O case as it is the most well studied molecule currently with a large variety of line list sources and three different broadening molecules available to test. Further, H$_2$O is one of the best measured molecules in exoplanet atmospheres to date. We anticipate expanding this work to cover other molecules such as CO with further broadening agents as it is another molecule of interest, particularly in very high resolution Doppler spectroscopy of hot exoplanetary atmospheres \citep[e.g.][]{2013ApJ...764..182S}.

However, despite these sources there is a lack of both experimental and theoretical high-temperature data on H$_2$ broadening parameters for many molecules of interest for exoplanetary atmospheres, e.g. CH$_4$, CO$_2$, NH$_3$, C$_2$H$_2$, HCN, TiO, CO, etc \citep{madhu2012}. Currently most molecules that are available with pressure broadening information have only self broadening and air broadening parameters. This is beginning to be addressed and more H$_2$ broadening parameters are becoming available \cite{Wilzewski2016193}.

There has been a great deal of work in making these parameters accurate and accessible which is made difficult by the sheer volume of lines for which broadening parameters are needed. Databases such as HITEMP, HITRAN and GEISA give pressure broadening values for each transition for each molecule. Where values are not available we are forced to turn to what is available in the literature from experiments and from other line lists. Where detailed line-by-line pressure broadening values are not available for a molecule we use the mean of pressure broadening values available from other sources. While this approach is not ideal it is currently the only option in some cases. For example in this work the YT10to10 line list for CH$_4$ is highly complete and a useful asset to modelling of exoplanetary atmospheres, however pressure broadening values have not been calculated. An alternative HITRAN list has pressure broadening components but only a small fraction of the molecular transitions are covered and the list is only appropriate for room temperature applications. We will discuss the consequences of this method of taking the average broadening in section~\ref{meanapproach} and the impact using a mean broadening value has on the final cross section. Table \ref{tab:allmols} also shows which molecules have detailed broadening available and which only have mean broadening. For each case where detailed broadening parameters are available a mean case has also been investigated in order to make a comparison.

\label{voigtvsgauss}

\section{Generating Cross-Sections}

The combination of molecular line parameters and broadening profiles discussed above allows us to compute the molecular absorption cross sections which form inputs to exoplanetary atmospheric models. We compute molecular cross sections over a wide range of pressures ($P$) and temperatures ($T$) relevant for exoplanetary atmospheres. Our grid in $P$-$T$ space is shown in Table~\ref{grid}. For each point in $P$-$T$ each cross section will usually be calculated multiple times using different broadening parameter or source. If a cross section was required between these points interpolation could be used on the cross sections between the nearest pressure and temperature values as outlined in \cite{2013Icar..226.1673H}. In this section, we describe the procedure we use to compute cross sections. 

\subsection{Line Intensities and Partition Functions}

Generation of cross sections requires the intensity of each transition to be accurately calculated. Most line list databases give Einstein coefficients for each transition with degeneracies and energies for each state. These can be converted into line intensities as \citep{2013JQSRT.130....4R}: 

\small
\begin{equation}
  \label{Intens1}
  S_{i,j}(T_{ref})=\frac{A_{i,j}}{8\pi c \nu_{i,j}^2Q(T_{ref})}g_ie^{-hcE_j/k_BT_{ref}}(1-e^{-hc\nu_{i,j}/k_BT_{ref}})
\end{equation}
\normalsize

where $A_{i,j}$ is the Einstein coefficient for spontaneous emission for the transition between states $i$ and $j$, $g_i$ is the upper state degeneracy, $E_j$ is the lower state energy in cm$^{-1}$ and $\nu_{i,j}$ is the transition frequency between $i$ and $j$, also in cm$^{-1}$ and finally $h$ is Planck's constant. Here, $Q(T_{ref})$ is the partition function at the required reference temperature. When an intensity is given at a reference temperature, usually 296 K, it can be converted to an intensity at any temperature using

\scriptsize
\begin{equation}
  \label{Intens2}
  S_{i,j}(T)=S_{i,j}(T_{ref})\frac{Q(T_{ref})}{Q(T)}\frac{exp(-hcE_j/k_BT)}{exp(-hcE_j/k_BT_{ref})} \frac{[1-exp(-hc\nu_{i,j}/k_BT)]}{[1-exp(-hc\nu_{i,j}/k_BT_{ref})]}.
\end{equation}
\normalsize
This gives the intensity of a transition in units of cm$^{-1}$/(molecule cm$^{-2}$). 

As evident from Eqs. \ref{Intens1} and \ref{Intens2} the partition function scales the line intensities. The partition function gives a measure of how many of the molecules of a gas are in the ground state compared with all other states. This ratio of state populations increases with temperature as it becomes more likely to find particles in higher energy states. This is intrinsically linked to the energy of each transition which makes the partition function unique for each molecule. The partition function is given by 
\begin{equation}
  Q(T)=\sum_jg_je^{-E_j/k_BT}
\end{equation}

where $g_j$ is the lower state degeneracy and $E_j$ is the lower state energy as described in Eq. \ref{Intens1}. However, to calculate a partition function this way the spectral information of the molecule must be complete. Missing transitions result in an inaccurate partition function which will not be representative, especially at high temperatures. Information on how partition functions are calculated for ExoMol can be found in \cite{2012MNRAS.425...21T}. Errors in the partition function may cause discrepancies that are small when considering individual cross sections at particular temperatures but may impact significantly when considering full atmospheric models. Models of atmospheres involve temperature-pressure profiles with respect to altitude and so require cross sections at many different temperatures to be included in computing the emergent spectrum. This will cause any errors in the cross section from the partition function to become compounded by many layers of an inhomogeneous atmosphere.

In the present work, partition functions have either been adopted from existing databases or computed using the TIPS code \citep{Fischer2003401}. The ExoMol database provides partition functions for use with each molecular line list. The HITRAN database uses the TIPS code as discussed in \cite{Fischer2003401} to create partition functions between 70K-3000K. There can be some discrepancies between partition functions from different sources, particularly at high temperatures. Some are only available in a small range of temperatures. Where this is the case we use extrapolation to find the values at higher temperatures. In our case this only affected our highest temperature point of T=3500K. 

Using a line list at a temperature greater than that recommended by the source is possible by using the correct partition function. However, while at low temperatures many transitions have low enough intensity to be discounted without effecting the final result, as temperature increases these low intensity lines begin to contribute more noticeably. Because of this any low transitions that are missed, (either by an intensity cut off that is too low or from being omitted in the source,) can cause errors in the cross section if molecular line lists are used above their recommended temperatures. These temperatures are given in the last column of Table \ref{tab:allmols}. Here we see with the exception of NH$_3$ and CH$_4$ the ExoMol and HITEMP line lists provide coverage up to and beyond the temperatures used in this work. The HITRAN sources, used primarily for earth applications, are only valid at room temperature. Due to their small number of transitions, (shown in Table \ref{sources}) HITRAN line lists are unreliable at high temperatures even when used  with the correct partition function. Nevertheless, these have been included in this work for two reasons; firstly to investigate the dependence of completeness of line list sources on the accuracy of cross sections and, secondly, to include useful molecules that are not covered by any other source.

\begin{center}
  \begin{table}
    \begin{tabular}{|cccccccc|}
      \hline
      T (K) & 300 & 400 & 500 & 600 & 700 & 800 & 900 \\
      - & 1000  & 1200 & 1400 & 1600 & 1800 &2000&2500 \\
      - & 3000 & 3500 &&&&& \\
      \hline
      P (atm) & 0.0001 & 0.001 & 0.01 & 0.1 & 1 & 10 & 100 \\
      \hline
    \end{tabular}
    \caption{Pressure-Temperature grid for cross sections. Values between these points could be interpolated by the user to allow a finer grid. These points have been chosen as they represent observable temperatures and pressures of currently known exoplanets.}
    \label{grid}
  \end{table}
\end{center}

\begin{table}
  \centering
  \begin{tabular}{c c}
    \hline 
    Wavenumber Range [cm$^{-1}$] & Grid spacing $\Delta\nu$ [cm$^{-1}$] \\
    \hline
    10-100 & 10$^{-5}$ \\
    100-1000 & 10$^{-4}$ \\
    1000-10000 & 10$^{-3}$ \\
    10000-30000 & 10$^{-2}$ \\
  \end{tabular}
  \caption{Table from \citep{2013Icar..226.1673H} giving the staggered spacing of the grid for used for line mapping. Here we present an alternative adaptive grid spacing which is given in Equation \ref{eq:grid}.}
  \label{tab:hillgrid}
\end{table}

\subsection{Cross-Sections from Line Intensities}
\label{sec:crosssections}
Cross sections are derived from line intensities by first broadening with the appropriate profile. This is followed by binning the resultant cross sections to the desired spectral resolution. This is a general approach followed by several recent studies \citep[e.g.][]{2013Icar..226.1673H}, albeit with minor differences in implementation. Here we discuss our implementation. The differences from other works are discussed in section~\ref{sec:optimal_grid}. For a given temperature ($T$) and pressure ($P$), computing the cross sections from line intensities involves three steps as follows. 

 Firstly, the Voigt profile ($f_v$) is computed at a high resolution in order to accurately evaluate each individual line profile as described in section~\ref{numericalmethod}. The spacing of this fine grid, here referred to as the `sub-grid', is given as 
 
\begin{equation}
\label{eq:grid}
  \Delta \nu=\frac{\gamma_{v}(\nu=500,T,P,m)}{6}
\end{equation}  

where, $\gamma_{v}(\nu=500,T,P,m)$ is the HWHM of a Voigt profile at $\nu = 500$ cm$^{-1}$. We find this prescription, which samples each Voigt width with 6 points, to provide the requisite resolution and accuracy with optimal computational speed, as discussed in more detail in section~\ref{sec:optimal_grid}. This very fine sampling gives an accurate representation the Voigt function across the whole wavenumber space. This sub-grid spacing is a function of $T$, $P$ and molecular mass making it specific to the molecule concerned.

Secondly, for a given spectral line, the cross section is computed at each point on the sub-grid described above. The cross section $\sigma (\nu)$ of a transition between states $i$ and $j$ at a certain pressure ($P$) and temperature ($T$) is given by 
\begin{eqnarray}
  \sigma_{i,j,P,T} (\nu) &=& S_{i,j,P,T}\frac{f_v(\nu)}{\int_{-\infty}^{\infty}f_v(\nu) \;d\nu} \nonumber \\ 
  &\approx& S_{i,j,P,T}\frac{f_v(\nu)}{\int_{\nu_{i,j}-\frac{\Delta\nu_c}{2}}^{\nu_{i,j}+\frac{\Delta\nu_c}{2}}\;f_v(\nu) \;d\nu}
  \label{eq:cross}
\end{eqnarray} 
in units of cm$^{-1}$/molecule where, $S_{i,j,P,T}$ is the line intensity and $f_v(\nu)$ is the Voigt function with broadening parameters corresponding to the line at the given P and T. $\nu_{i,j}$ is the wave number of the line centre and $\Delta\nu_c$ is the extent of the profile to which the line wings are evaluated. For a given line, we use a $\Delta\nu_c$ value of 500 Voigt widths, (250 around the line centroid,) which we find to be optimal, as discussed in section~\ref{sec:optimal_grid}. The integral is evaluated up to this cut off and normalises the profile. Evaluating up to this cut off effectively folds in the intensity from the missing wings that are not evaluated back into the profile, ensuring no intensity is missed.

Finally, the high-resolution cross section profile computed for each line as described above is binned to a final cross section grid with a coarser spacing for saving on storage space. The final cross section grid spacing is still high resolution at $10^{-2}$ cm$^{-1}$ which corresponds to a resolution of R=100,000 or greater depending on wavelength. For instance, at a wavelength of 1 $\mu$m this spacing corresponds to a spectral resolution (R = $\lambda/d\lambda$ = $\nu/d\nu$) of $10^6$. When a lower resolution cross section grid is desired, this high-resolution grid can be binned down further in frequency-space or wavelength-space as required. For example, for a given resolution R the grid points in wavelength space can be determined following $R = \lambda/d\lambda$ which will give a non-linear grid in $\lambda$. The mid points between adjacent bins are selected and all values within these bounds are averaged giving the binned down contribution at each wavelength on the grid. 

In this work as, in discussed in \cite{2013JQSRT.130....4R} and other works, we apply a cut in intensity to only evaluate the significant lines. This provides a reduction in computation time with minimal effect on  accuracy as very low intensity lines contribute little to the final cross section even at high resolution. We apply a cut off at $10^{-30}$ cm$^{-1}$/(molecule cm$^{-2}$) in intensity across all line lists apart from the BYTe and YT10to10 line lists for NH$_3$ and CH$_4$ respectively. Due to their large size the cut off was increased to $10^{-26}$ cm$^{-1}$/molecule cm$^{-1}$. We find that as both these molecules have many complex transitions our results are unaffected by omitting these low lines across all resolutions covered. Unlike the HITRAN database we do not scale our cross sections by any relative abundances of isotopes based on their terrestrial measurements. This gives each cross section with the molecule at 100\% abundance.

The complete list of molecules used in this work including all sources for data is given in Table~\ref{tab:allmols}. Many different line list sources were chosen, particularly for water, for comparison to investigate how completeness effects the resulting cross section.

\begin{table}
  \begin{minipage}{5cm}
    \centering
    \begin{tabular}{cccc}
      \hline
      \textbf{Molecule} & \textbf{Source} & \textbf{Broadening} & \textbf{Max T (K)}   \\ 
      &  & \textbf{Agent} &   \\\hline
      H$_2$O & BT2\footnote{\cite{2006MNRAS.368.1087B}} & Self, Air&  3000K \\
      H$_2$O & HITEMP\footnote{\label{foot:hitemp}\cite{2010JQSRT.111.2139R}} & Self, Air &  4000K \\
      H$_2$O & HITRAN\footnote{\label{foot:hitran}\cite{2013JQSRT.130....4R}} & Self, Air & 296K \\
      H$_2$O & PS (1997)\footnote{\cite{PS}} & H$_2$ & - \\
      CO$_2$ & HITEMP\footref{foot:hitemp} & Self, Air  & 4000K \\
      CO$_2$ & HITRAN\footref{foot:hitran} & Self, Air & 296K \\
      CO & HITEMP\footref{foot:hitemp} & Self, Air &  4000K \\
      CO & HITRAN\footref{foot:hitran} & Self, Air &  296K \\
      OH & HITEMP\footref{foot:hitemp} & Self, Air &  4000K \\
      OH & HITRAN\footref{foot:hitran} & Self, Air &  296K \\
      NO & HITEMP\footref{foot:hitemp} & Self, Air &  4000K \\
      NO & HITRAN\footref{foot:hitran} & Self, Air &  296K \\
      CH$_4$ & YT10to10\footnote{\cite{2014MNRAS.440.1649Y}} & Self, Air & 2000K \\
      CH$_4$ & HITRAN\footref{foot:hitran} & Self, Air & 296K \\
      NH$_3$ & BYTe\footnote{\cite{2011MNRAS.413.1828Y}} & Self, Air & 1600K \\
      NH$_3$ & HITRAN\footref{foot:hitran} & Self, Air & 296K \\
      HCN & Harris\footnote{\cite{2006MNRAS.367..400H}} & Self, Air &  4000K \\
      HCN & HITRAN\footref{foot:hitran} & Self, Air &  296K \\
      C$_2$H$_2$ & HITRAN\footref{foot:hitran} & Self, Air &  296K \\
      \hline
    \end{tabular}
  \end{minipage}
  \caption{Molecules used to generate cross section database and all line list sources, including which broadening agents are given. (Note HITRAN is recommended at room temperature of 296K.) Here we choose many different sources to compare the effect that completion has on our final cross sections.}
  \label{tab:allmols}
\end{table}

\section{Optimal Resolution and Cut off of Broadening Profile} 
\label{sec:optimal_grid}

It is important to consider an appropriately high resolution and extent of the line profile in order to accurately sample the contribution from the profile. Both these properties also influence the computational cost. Therefore, it is desirable to adopt optimal values for each of the properties which facilitate both a high enough accuracy and a reasonable computation time. Several recent studies have adopted different prescriptions for these parameters in the particular context of molecular cross sections for exoplanetary applications \citep[e.g.][]{2013Icar..226.1673H,2015arXiv150303806G}. In this section, we systematically investigate the effect of both the profile grid resolution as well as the extent of the profile wings on the cross sections in an attempt to determine optimal values for these parameters. 

\subsection{Effect of Profile Grid Resolution} 

In the present work we use a grid resolution that is adaptive with the equivalent width of the Voigt profile for a given line profile as given in Eq \ref{eq:voigtwidth}. This approach allows for optimal computational time while ensuring high accuracy of the cross sections. In this formulation, the grid in frequency space on which we evaluate the Voigt profile, here referred to as the `sub-grid' , is defined by Eq. \ref{eq:grid}. In this work we use a minimum sampling of 6 points per Voigt width which was found to be sufficiently accurate based on the investigations carried out below. Evaluating this spacing at $\nu=$500 cm$^{-1}$ gives a conservative estimate of the width of a Voigt profile where the Gaussian component is narrowest. (See Equation \ref{eq:gausswidth}.) This also corresponds to a wavelength of 20 $\mu$m which is the longest wavelength of interest in the present work and the upper wavelength limit of the infrared observations of exoplanetary atmospheres for instruments such as MIRI on JWST. The necessary grid spacing can become very wide at high pressures where the equivalent width of the Voigt profile becomes large. Therefore, we place an upper-limit of 0.01 cm$^{-1}$ on the grid spacing.  
When evaluating Voigt profiles on this grid the range up to which the profile is calculated is given by $\Delta\nu_c$, as shown in Eq~\ref{eq:cross}, and discussed in detail in section \ref{EvaluationWidth}. 

\begin{figure}
  \centering
  \includegraphics[scale=0.4]{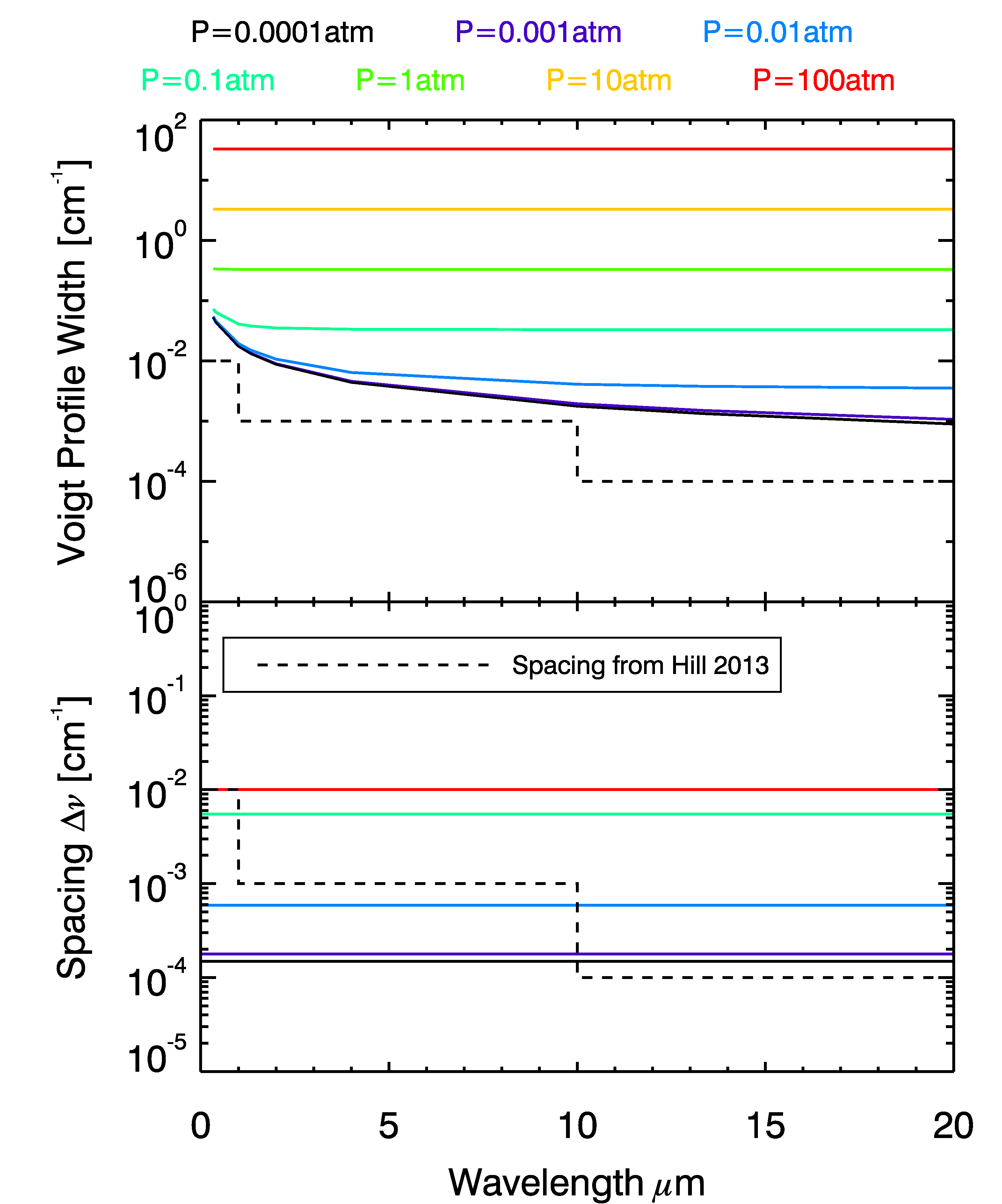}
  \caption{Comparison of grid spacing in our adaptive grid with that in the staggered grid from Table~\ref{tab:hillgrid} of \citet{2013Icar..226.1673H} for different pressures. Top: The coloured lines show the Voigt profile widths at different pressures and a representative temperature of 1000 K. The staggered grid spacing is shown in dashed line for reference, demonstrating that the grid spacing is well below the Voigt profile width for pressures down to 10$^{-4}$ atm. Bottom: The coloured lines show the adaptive grid spacing we use. For high pressures ($\geq$1 atm), the minimum spacing is fixed at 10$^{-2}$ cm$^{-1}$ as profiles become very broad. While our grid spacing is coarser than the staggered grid spacing for high pressures the resultant effect on the cross sections is small, as shown in Fig.~\ref{VoigtParams2}. Note that a single value of broadening width and temperature scaling has been adopted for this figure. }
  \label{VoigtParams}
\end{figure}

Our spacing is coarser than that used in some previous studies but is optimised for computational time and accuracy in computing cross sections. For example, the spacing used in \cite{2013Icar..226.1673H}, as shown in Table~\ref{tab:hillgrid} and referred to here as a staggered grid, is finer than the grid employed here. This is in part due to work from \cite{2013Icar..226.1673H} concerning only thermal broadening where their profile is Gaussian and much narrower requiring finer grid spacing to normalise. However using our coarser, adaptive grid does not cause significant errors in the final cross sections. Figure~\ref{VoigtParams} shows a comparison between the spacing of our adaptive grid and the staggered grid for a representative temperature of 1000 K. It can be seen that both grids are much finer than the Voigt widths for each pressure case, (upper panel). While a fine grid spacing gives highly accurate profiles using the staggered grid can lead to unnecessarily high resolution, especially in the limit of high pressures where the profiles become inherently very broad. This can be computationally expensive particularly for high pressures. The grid we propose in Eq. \ref{eq:grid}, referred to here as an adaptive grid as its spacing with changes in pressure, uses fewer points at higher pressures to overcome this problem while preserving the accuracy. 

Ensuring the profile is evaluated accurately is imperative to achieve an accurate normalisation and neither under nor over represent the line intensity contribution at each grid point. For normalising the profiles we compute the area under the curve using a simple trapezium rule. When a profile is only sparsely evaluated this approach will tend to a greater area estimation, producing a lower intensity contribution from each line profile after normalisation. This results in a small percentage of ``missing'' intensity. To test the validity of our approach we analyse this amount of missing intensity in a single profile when calculated using the adaptive grid compared with the finer, staggered grid from \cite{2013Icar..226.1673H}. This is done for each T and P point.

\begin{figure}
  \centering
  \includegraphics[scale=0.3]{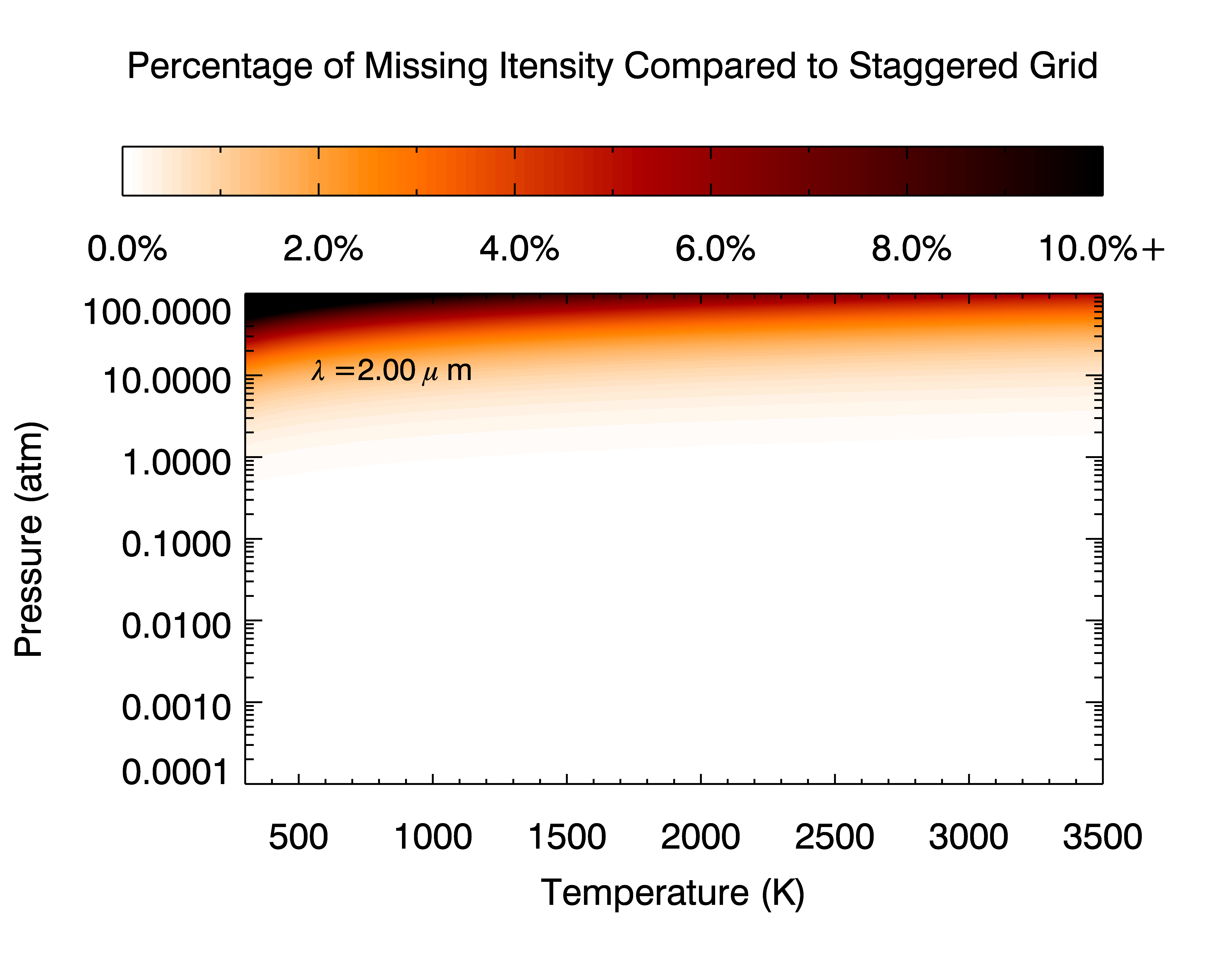}
  \caption{Comparison of adaptive grid with staggered grid from Table~\ref{tab:hillgrid} . Voigt profiles are mapped to a fine grid from equation \ref{eq:grid}. The integrated area under the profile for the adaptive case is compared with that for the staggered case to produce this result. Only pressures of $\sim$100 atm are affected significantly by the resolution of the adaptive grid beyond a few percent. Below 1 atm the difference is $\approx$ 0.2\% or less.}
  \label{VoigtParams2}
\end{figure}

In order to test our adaptive grid and how well it approximates the Voigt profile it is compared with the staggered grid using a wide cut off of $\Delta\nu_c$=100 cm$^{-1}$ wavenumbers for each pressure and temperature case. Here we take one profile per T and P point only and map to each grid. The profiles are generated in the same way as in section \ref{numericalmethod}. Any difference in the profiles will effect the integrated area, which is used for normalisation. We consider the staggered grid to be high enough resolution in all cases that it will produce an accurate area estimation. The comparisons have been conducted over a range of wavelengths but here we select a representative wavelength of $\lambda=$2$\mu$m for illustration. Figure~\ref{VoigtParams2} shows the results of this comparison. The difference in the final output of these grids is very small with $\lesssim$0.2\% of intensity missing  at P$<$1 atm at all temperatures. At P=10 atm a maximum of 2\% of the original intensity is missed at low temperatures. The largest differences found are $\sim$10\% for pressures of 100 atm, but given that such high pressures are not directly observable for exoplanetary atmospheres the corresponding differences are less of a concern. We find that in such cases the cut off chosen $\Delta\nu_c$ is too narrow for the extreme case of P=100 atm. For the 10$^{-4}$ to 10 atm pressure range we find the adaptive grid to be very accurate, particularly after binning to the final output grid with a spacing of 0.01 cm$^{-1}$ which corresponds to a spectral resolution of R=10$^5$-10$^6$ at $\lambda <$10 $\mu$m. 

\subsection{Effect of Profile Evaluation Width}
\label{EvaluationWidth}

In order to accurately account for the contribution of the line wings the broadening profile must be evaluated over a wide enough range centred on the line centre, referred to here as a cut off value ($\Delta\nu_c$), as discussed in section \ref{sec:crosssections}. This cut off value determines the extent of evaluation of the profile as well as its normalisation as described in Eq.~\ref{eq:cross}. The choice of $\Delta\nu_c$ has already been noted in the field as an important factor in computing cross sections \citep{2007ApJS..168..140S,2015arXiv150303806G}. A common approach is to apply a cut off in wavenumber, with values ranging between 10 and 100 cm$^{-1}$, especially for high pressures as discussed by \cite{2007ApJS..168..140S}. Another approach is to take a number of Lorentzian widths from the centroid such as in \cite{2015arXiv150303806G}, however this does not take into account the full width of the profile after its convolution with the Gaussian component. We employ a cut off in multiples of Voigt widths, given by Eq~\ref{eq:grid} and implemented in Eq~\ref{eq:cross}. The $\Delta\nu_c$ in this approach adapts with both the Lorentzian and Gaussian components ensuring that the wings of the profile are accounted for in an adaptive manner depending on the broadening conditions. 

Based on our investigation we find 500 Voigt widths to be sufficient for current applications including JWST-like resolutions and VLT applications with small uncertainties when using the standard Voigt profile. 

Several cut off values have been investigated to establish the optimal balance between accuracy and computational time. As discussed above we use multiples of Voigt widths to establish a cut off that adapts to the specific profile.  When using a short cut off lower intensity lines are underestimated by many orders of magnitude due to the lack of additional intensity from the wings of high intensity neighbours. This leads to the continuum being poorly approximated by short cut offs. When the cut off is increased we maintain the same normalised area within the profile. Because of this the profile height becomes slightly shorter as the cut off becomes wider. Due to this effect we find that at low cut off values the cores of very strong lines are slightly over estimated, (by approximately 0.2\%) at the native spacing of the grid. The effect of this slight overestimation will reduce drastically with resolution.

The underestimation in the profile wings is ~10\% for H$_2$O at the native spacing. However this underestimation is confined to the lowest intensity transitions with high intensity neighbours which are by their nature confined within high intensity features. As such an underestimation in these points is of less consequence to most applications.

Due to the profiles being very broad at high pressures where the Voigt profile width is beyond 0.01 cm$^{-1}$ the spacing becomes fixed to 0.01 cm$^{-1}$ wavenumbers. This also fixes our evaluation width to 60 cm$^{-1}$ wavenumbers around the centroid. For high pressures, (P$>$ 10 atm), this is an underestimate, however such pressures are less useful for atmospheric modelling in exoplanets.

\begin{figure}
  \centering
  \includegraphics[scale=0.36]{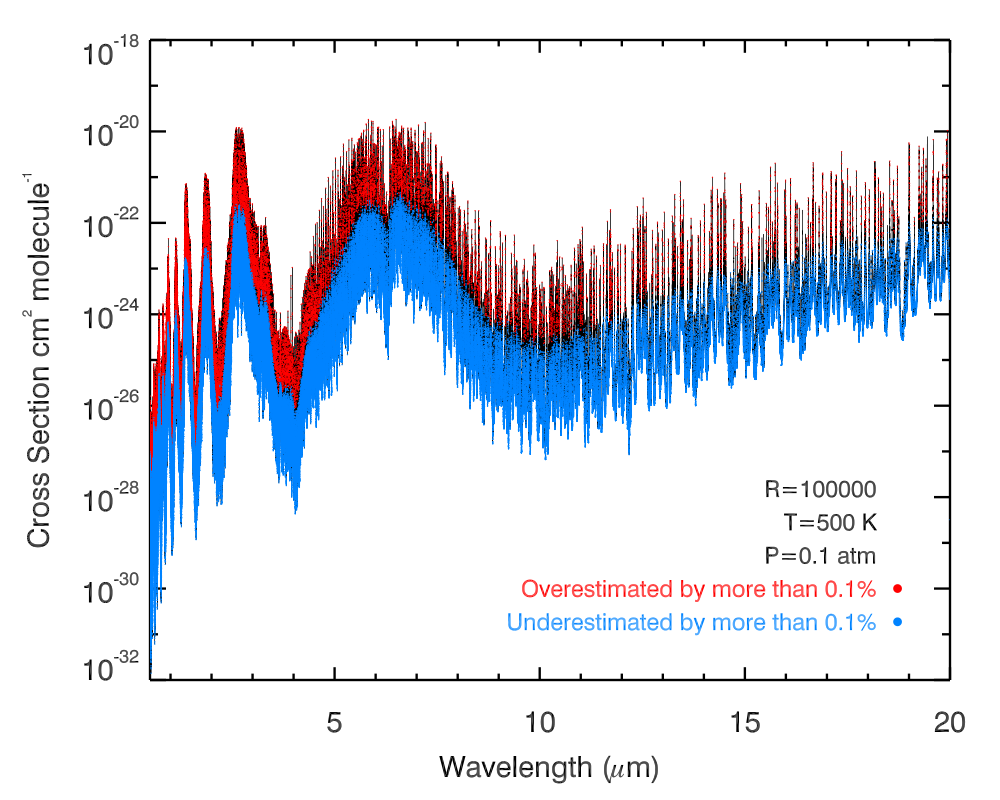}
  \caption{Comparison of cross sections obtained at the highest resolution using profile evaluation width ($\Delta\nu_c$) of 500 Voigt widths versus 10,000 Voigt widths for H$_2$O at 500K and 0.1 atm. Overestimated  and underestimated points are shown in red and blue, respectively. We see there are some points overestimated by more than 0.1\%, however no points are overestimated by more than 1\%. The underestimations average ~10\% at this high resolution.}
  \label{percentagevoigt}
\end{figure}

\subsection{Comparison of Evaluation Widths}
\label{sec:evalwidth}
The wings of the profile will have an affect on any neighbouring lines. If wings are not evaluated out to a large enough separation the continuum for neighbouring lines will be underestimated. This is particularly important in the case of line lists and cross sections as intensities span many orders of magnitude and so line wings from high intensity transitions can be comparable to the peaks of low intensity neighbours. However the cut off also affects the evaluation of the profile. A cut off that is too close to the centroid will provide poor normalisation and will miss some of the intensity contribution. 

The cut off value $\Delta\nu_c$ is fixed at 500 Voigt widths around the centroid. This is designed so that the evaluation width adapts to both the Gaussian and Lorentzian profiles which is particularly important at extremes of pressure or temperature. This gives 250 Voigt widths around the centroid in each direction. $\Delta\nu_c$ is increased to 1000 Voigt widths around the centroid at pressures of 1 atm and above. To establish the difference between this method and others in the field we undertake the following comparisons. Firstly we take a single Voigt profile on our fine, adaptive grid given by Eq. \ref{eq:grid} and $\Delta\nu_c$ of 500 Lorentzian widths as taken from \cite{2015arXiv150303806G}. The Voigt profile is calculated in the same way as given in section \ref{numericalmethod}. This is then compared with a single Voigt profile on the same grid with a cut off at 500 Voigt widths, our value of $\Delta\nu_c$. The profiles are left unnormalised. The missing area from the Lorentzian cut off when compared with the method we present is then evaluated as a percentage of the total profile area. 

\begin{figure}
  \centering
  \includegraphics[scale=0.3]{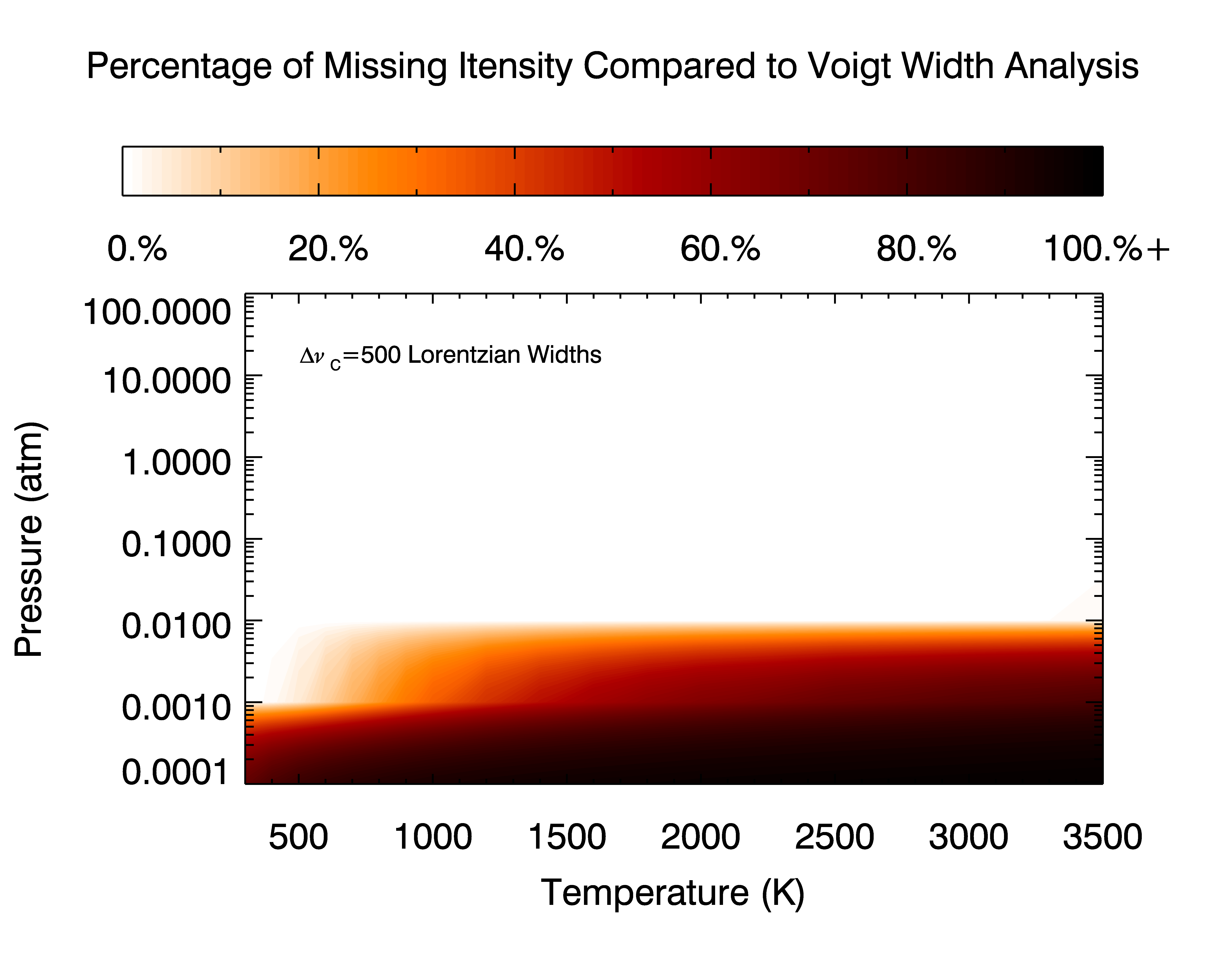}
  \caption{Comparison of the Voigt width cut off from Eq \ref{eq:grid} with $\Delta\nu_c$ set at 500 Lorentzian widths. For pressures of 0.001 atm and below we see that missing intensity can be as high as $\sim$50\% leading to a loss of information at these crucial low pressures.}
  \label{VoigtParams3}
\end{figure}

The percentage of missing intensity from this comparison is shown in Figure~\ref{VoigtParams3}. It can be seen from the Lorentzian and Voigt width equations that the Voigt width should always cover the same or more wavenumbers and so the Lorentzian width cut off has a smaller integrated area than our approach.

The Lorentzian width evaluation fails at low pressures as the thermal Gaussian component becomes much stronger at lower pressures. These low pressures are crucial to understanding the upper atmospheres and are the most likely to effect observations. Any intensity missed in the initial evaluation of the profile then leads to inaccuracies to each individual profile which can then be further compounded by binning several such transitions to the final output grid. An underestimation of the true area under the profile results in an incorrect normalisation and an overestimation of the cross section value for each transition. We see in Figure \ref{VoigtParams3} this overestimation can be ~50\% for pressures of 0.001 atm.

\begin{figure}
  \centering
  \includegraphics[scale=0.3]{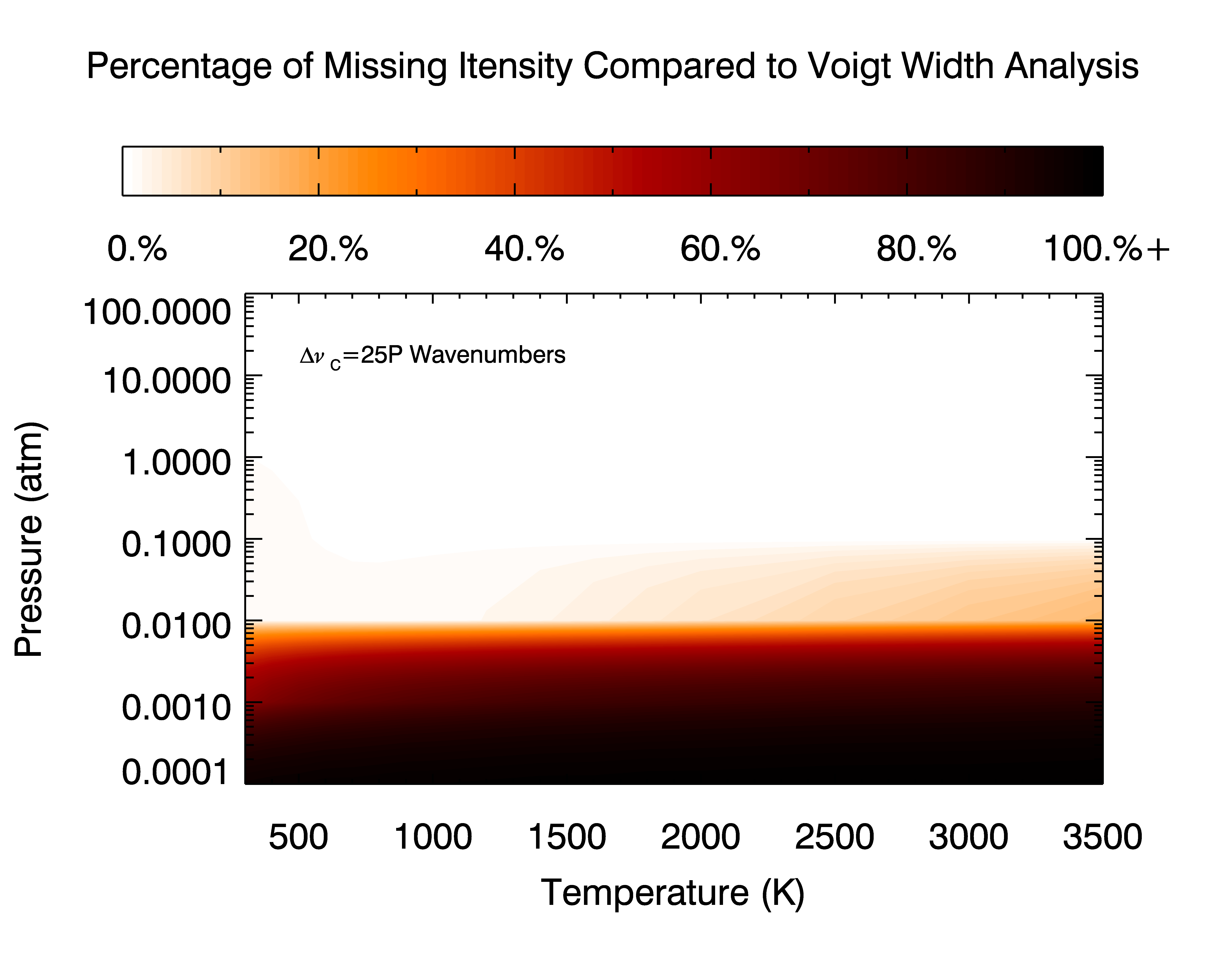}
  \caption{Comparison of the Voigt width cut off from Eq \ref{eq:grid} with a cut off set at 25P wavenumbers (up to a maximum of 100cm$^{-1}$) around the centroid. For pressures of 0.001 atm and below the differences are $\sim$50\% or greater.}
  \label{VoigtParams4}
\end{figure}

Another approach, taken from \cite{2007ApJS..168..140S}, is to apply $\Delta\nu_c$ of a certain number of wavenumbers from the centroid. \cite{2007ApJS..168..140S} suggest $\Delta\nu_c$ = min$(25 P, 100)$ cm$^{-1}$, where P is pressure in atm. This approach also adapts in pressure. However we find this also provides an underestimation of the area under the Voigt profile at low pressures as shown in Figure \ref{VoigtParams4}. 

Based on these parameters we find our combination of a sparse, adaptive grid with a wide cut off to be an improvement on current methods in the field in terms of both accuracy and reduction in number of points which require evaluation. We also find our method to give sufficiently accurate profile estimates for the final output grid spacing of 0.01 cm$^{-1}$ and the instrument resolutions given here in later sections.

\subsection{Sub-Lorentzian Shapes}

In this work, as discussed in Section \ref{sec:broadening}, we use only the standard Voigt profile without modification to account for changes in the shape from other processes, (e.g. line-mixing and collisionally induced velocity changes) such as those discussed in \citet{2012RSPTA.370.2495N}. \citet{1991JGR....9620859E} and \citet{Birnbaum1979597} discuss specifically a sub-Lorentzian shape for pressure broadening where the far line wings are modified by an exponential, reducing their contribution. Currently there are many estimates of the distance from the line centroid where this occurs ranging from 1-30cm$^{-1}$. This sub-Lorentzian shape at the far line wings can be difficult to estimate and contributions to the profile at large separations are not well understood.  Changes to the broadening shape used here could be made in the future to investigate whether a change in profile shape significantly alters results.Molecular transitions vary greatly in intensity and this may have a greater effect in the wavelength regions where there is a sharp drop in line intensity ('window regions'). For some molecules with less complex structure, such as CO, these can be very sharp changes. In such cases, having no neighbouring lines, a sub-Lorentzian may have some effect on the edges of these regions.

Other approaches have included a simple cut off at a given wavenumber from the centroid of the line which would be very similar to the tests performed in Section \ref{sec:evalwidth}. We find that changes in this cut off cause small underestimations in the lower intensity transitions across wavelength. However we find our metric, discussed in Section \ref{measuring}, to be quite general even accounting for this effect. At most resolutions many transitions are binned together mitigating this problem and our metric uses a median over a wide band.

\section{Effect of Pressure Broadening on Cross sections} 

In this section we systematically investigate the dependence of molecular cross sections on the various parameters and assumptions involved in implementing pressure broadening with a given line list. Our goal here is to both quantify the uncertainties propagated into cross sections due to various choices involved in their generation as well as to define a quantitatively meaningful approach to make those choices. In order to pursue this here we focus on the H$_2$O molecule which has the most complete line list and broadening data currently available for exoplanetary atmospheres. We investigate the dependence of the cross sections on the following key factors: (a) pressure and temperature, (b) average versus line-by-line treatment of broadening parameters, (c) spectral resolution (i.e. binning) of the cross sections, and (d) broadening agent. Note that the effect of the parameters of the Voigt line profile, namely the profile resolution and extent of the line wings, were investigated in the previous sections and here we adopt the optimal values and methodology discussed in sections \ref{sec:crosssections} and \ref{sec:optimal_grid}. 


\subsection{Definition of Change Due to Broadening}
\label{measuring} 

In order to assess the difference that any given aspect of broadening makes to the cross sections we need to formally define the corresponding change quantitatively. Since the cross sections for any molecule can span many orders of magnitude over a given spectral range and features very drastically with wavelength defining a robust metric is challenging. For example, a simple metric such as a 'mean difference' across the entire spectral range available is often unreliable. On the other hand, focusing on lines with maximum error places undue emphasis on the lowest intensity lines which will see the highest fractional change but would be hard to observe. Conversely, focusing on the highest intensity lines is unreliable because they are not representative of the line population. Therefore, we use the median percentage difference across the entire spectral range as our metric of choice to quantify the change in cross sections due to any aspect of broadening. The median percentage change in cross section for each line is computed as 

\begin{equation}
\delta = {\rm Median}\Big\{\frac{\sigma-\sigma_0}{\sigma_0} \Big\}_\lambda\times100
\end{equation}

where $\sigma_0$ and $\sigma$ are the cross sections before and after incorporating the particular pressure broadening prescription, and the median is evaluated for cross sections computed over the entire wavelength range of interest, 0.5 - 20 $\mu$m. 


\subsection{Effect of Pressure and Temperature}

\label{PressureDifference}

\begin{figure}
 \centering
 \includegraphics[scale=0.5]{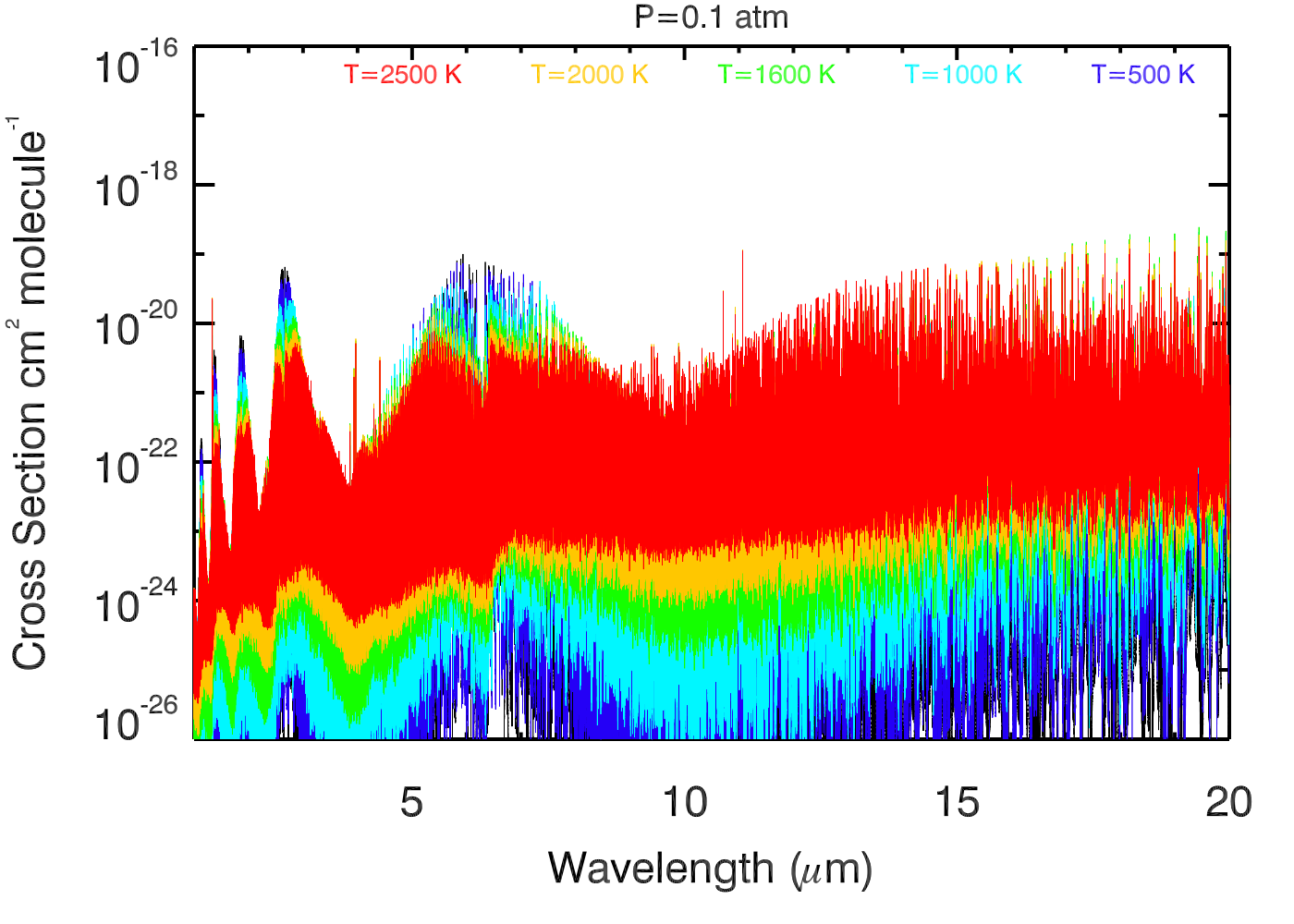}
 \includegraphics[scale=0.5]{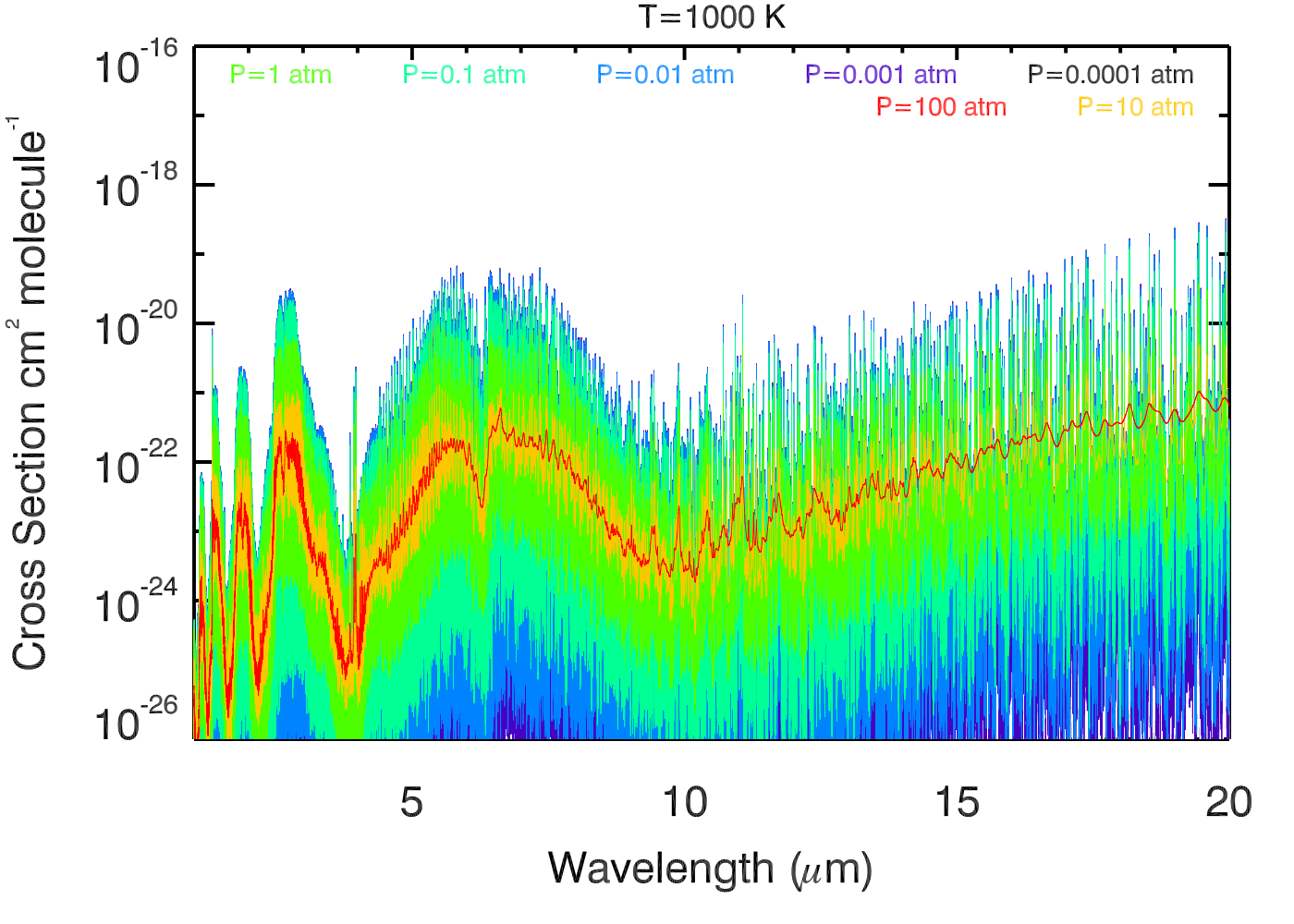}
 \caption{Cross sections for H$_2$O at various temperatures and pressures using air broadening at a grid spacing $\Delta\nu$ of 0.01 cm$^{-1}$ wavenumbers, the output grid of this database. This corresponds to a resolution up to an order of magnitude higher than R=10$^5$ below 10 $\mu$m.}
 \label{H2O_TempH2O_Pres}
\end{figure}

The effect of pressure broadening on cross sections as a function of pressure and temperature is shown in Figure \ref{H2O_TempH2O_Pres}. In the region of interest for exoplanet atmospheres, for pressure around 10$^{-4}$ to 1 atm and temperatures of 500 to 2000K) we find notable changes to the cross section as a function of resolution. For low resolutions (R $<$ 100), we find the differences introduced to cross sections from pressure broadening is $<$1\% across all P and T considered in this work. At resolutions of R=5000, similar to those expected from JWST, and representative hot Jupiter temperatures of T$\sim$1000K we find $\delta<$1\% at P=0.1 atm. At the highest resolution of R=10$^5$ we find that for temperatures of 1000K and pressures of 0.1 atm $\delta=$60\% for H$_2$O in an H$_2$ atmosphere. For lower temperatures of 500K this can increase dramatically giving $\delta$=1000\%. 

\begin{figure}
   \centering
   \includegraphics[scale=0.23]{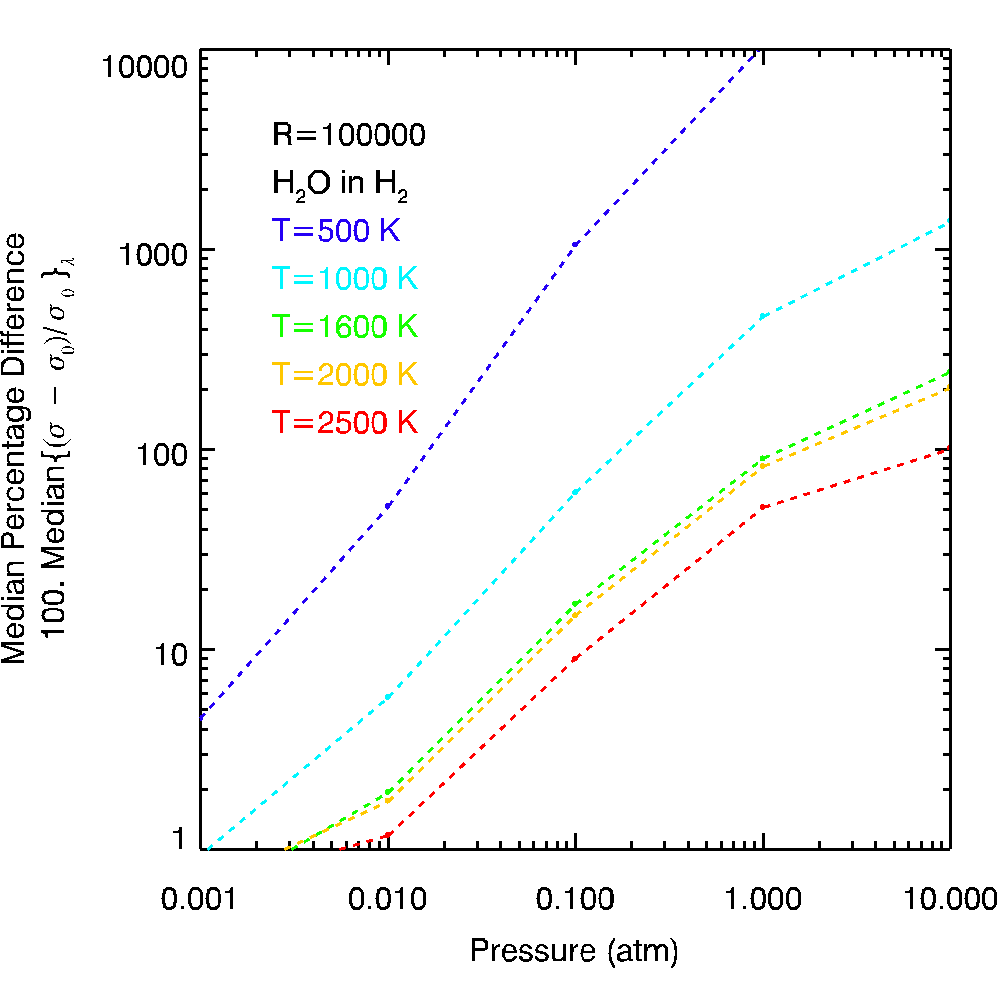}
   \caption{Median difference between cross sections derived with Gaussian-only broadening and Voigt broadening in the WFC3 band pass for H$_2$O at a resolution of R=100,000. Here we show the H$_2$ broadening case, though self broadening case gives the largest change to the cross section overall from pressure. For low temperatures the difference dramatically increases as the Gaussian component of the Voigt becomes narrower.}
   \label{fig:temppressres}
 \end{figure}

Figure \ref{ptregions} shows that at low temperatures (T$\lesssim$500K) the wide Lorentzian will have more of an effect than the Gaussian profile. From this we expect to see that cross sections, (and full atmospheric models,) at low temperatures are more effected by pressure broadening than those at high temperatures. Figure \ref{fig:temppressres} shows that, even before propagating through an atmospheric model, we can expect cool targets of $\sim$500K to be more affected by pressure broadening than hotter targets $\sim$1000K by 100\% or more at the highest resolutions. At resolution of R=$10^5$ we can expect the median difference between the two broadening cases, (thermal only and pressure and thermal,) for water to be 100\% beyond pressures of 0.1 atm at T=500K as shown in Figure \ref{fig:temppressres}. This would lead to uncertainties in molecular abundance of the a similar factor. The implication here is that cooler targets are likely to be more effected by pressure broadening. When observing such cool targets with high resolution instruments pressure broadening will potentially limit the precision on abundance measurements. These cooler targets are likely to have complex atmospheric structure and incorporate different chemistry in their upper atmosphere, even including broadening from more complex molecules than molecular hydrogen. However for cool targets there may be other factors obscuring observations such as clouds and hazes. 

In recent years, observational programs have largely focussed on hotter targets, mostly hot Jupiters, as their thick atmospheres and high temperatures lead to stronger spectral features and their short orbital periods make them easier to observe  than other targets. Currently observing lower-mass, cooler planets is proving difficult, however pressure broadening may affect us more in the future when observations become more sensitive to such planets. At temperatures of 1000K and beyond we still find an effect from pressure broadening though less strong. For these hotter targets we might expect pressure broadening to have an effect in the 0.1-1 atm pressure regime of around 10-100\%. As discussed in section \ref{sec:broadening_agent} this depends also on the broadening agent and we find that self broadening for water is much stronger than H$_2$.

\subsection{Line-by-line versus Mean Broadening Parameters} 
\label{meanapproach}

As currently there is a lack of line-by-line broadening parameters for several molecules it is often unavoidable to rely on sparse broadening data when computing pressure broadened cross sections, as discussed in section~\ref{sec:availability}. When only sparse data is available, i.e. broadening parameters are available only for a few lines, in this work representative values for the broadening parameters are chosen based on the available data and the mean applied across all the lines. We find that the difference to cross sections when using this method is up to $\sim$20\% at the output grid spacing of 0.01 cm$^{-1}$ for all pressures. In this section, we investigate the difference such an approach makes to the cross sections overall compared to cases where broadening parameters are available for all the lines. 

\begin{figure}
  \centering
  \includegraphics[scale=0.23]{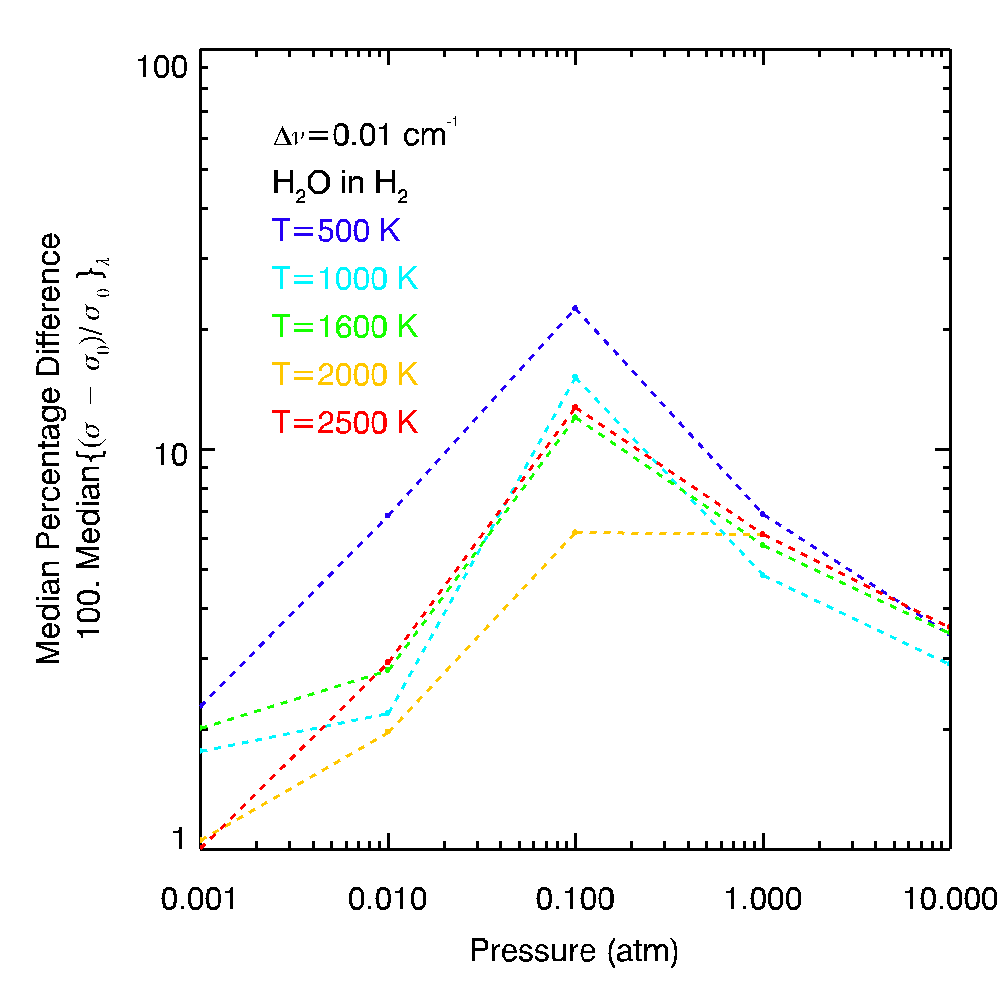}
  \caption{Effect of mean versus detailed pressure broadening. The curves show comparisons between cross sections obtained using mean values for all the line broadening parameters versus those obtained using  detailed line-by-line broadening parameters. These results are for the highest-resolution spacing of 0.01 cm$^{-1}$, corresponding to a resolution of 10$^5$-10$^6$ at $\lambda<$10 $\mu$m.}
  \label{fig:MeanCompare}
\end{figure}

 For purposes of demonstration, we use the latest line list of H$_2$O from the HITEMP database which includes line-by-line broadening parameters with air broadening. In one case, we calculate H$_2$O cross sections over a wide range of temperatures and pressures using detailed line-by-line values for the air-broadening parameters. In another case, we adopt constant values for the broadening parameters averaged over the entire line list and apply those values for broadening every line. The median percentage difference in the cross sections derived from the two cases over the entire line list at the native spacing of the line list with a grid spacing of 0.01 cm$^{-1}$ is shown in Figure \ref{fig:MeanCompare}. At this resolution, we find that using mean broadening values can result in cross sections that are inaccurate by up to 20\% for observable pressures ($\sim$0.1 bar) and low temperatures (T $\lesssim$500 K) where pressure broadening is strongest. For lower pressures and higher temperatures the effect is less pronounced in a `median' sense. For lower resolutions, the differences reduce. Therefore, when numerous lines are available to calculate representative average values for the broadening parameters as in the present case then the mean treatment of pressure broadening is a reasonable approximation to a detailed treatment for low resolution observations. 

 Figure \ref{fig:MeanCompare} shows this difference to decrease as pressure increases. In high pressure cases the broadening becomes wide enough such that many profiles begin to overlap. This effectively smooths the information and causes the the differences in profile shapes to be less distinguishable. From Figure \ref{H2O_TempH2O_Pres} this effect is more clear as we see at high pressures much of the information from the individual transitions is lost. This effect occurs at pressures greater than 0.1 atmospheres implying that for high pressures of P$\gtrsim$1 the detailed broadening parameters for each transition may not affect cross sections as much.

Contrary to the above scenario where we can construct a reasonable mean from many broadening parameters, detailed line-by-line broadening parameters are required to derive accurate cross sections, especially from upcoming high-precision and high-resolution observations. In the above case, while the median error across all the lines is low overall, individual lines with significant deviation from the mean broadening values can result in more than 100\% difference in cross sections which are relevant for interpreting high-resolution observations that rely on detecting specific lines \citep{snellen2010,2014Natur.509...63S}. Secondly, for several molecules relevant broadening values are available based on experimental data for only a few lines, the average of which may not be representative of the entire line population, leading to larger inaccuracies than found in the above example. Finally, recent and upcoming observations are already sensitive to cool and dense planetary and Brown Dwarf atmospheres \citep[e.g.][]{fraine2014,buenzli2015} with high-precision observations which necessitate accurate line-by-line broadening parameters.

Appendix \ref{LineListSources} shows all molecules that are available from various sources with how many lines each contains which here we use as a proxy for how complete a line list is, with those containing only a few thousands of lines being most unreliable. HITEMP, HITRAN and GEISA have line by line pressure broadening parameters generated but only for air and self broadening, with only HITEMP containing values for the high temperatures relevant for exoplanetary atmospheres. In order to address the issue of pressure broadening accurately in the future line lists will need to include broadening due to molecular hydrogen, relevant for giant planets, and be complete to high temperatures. 

\subsection{Effect of Spectral Resolution}

 \begin{figure}
   \centering
   \includegraphics[scale=0.5]{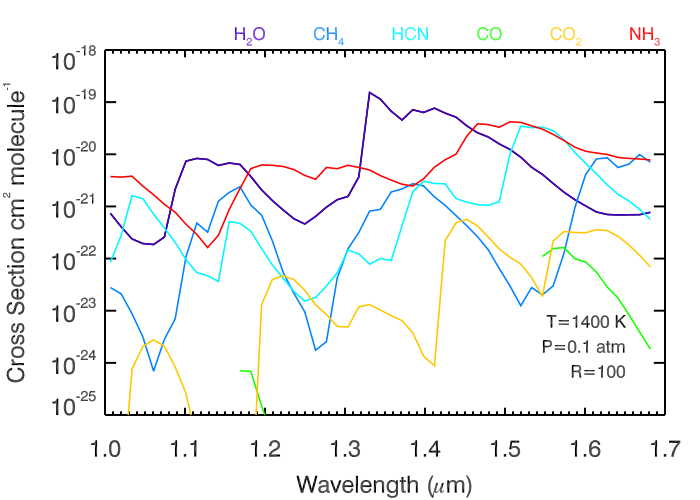}
   \includegraphics[scale=0.5]{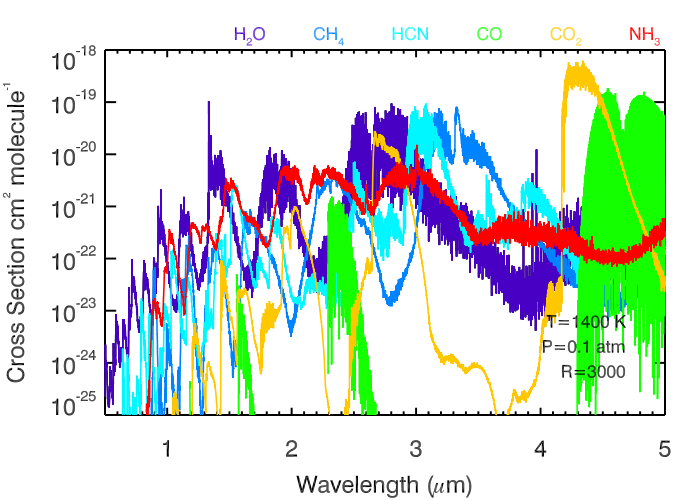}
   \caption{Comparison of molecular cross sections in the WFC3 G141 bandpass and the NIRSpec bandpass at HST-like and JWST-like resolutions, respectively. The cross sections are generated using air broadening for all the molecules. Degeneracies between the specific molecule, the abundance of the molecule and the temperature can be harder to break at lower resolutions, however JWST's improved wavelength and resolution will help break these degeneracies in future.}
   \label{molzoom}
 \end{figure}

One of the most important questions that can be answered in this work is how the difference created by pressure broadening to molecular cross sections is influenced by spectral resolution. As discussed earlier, we use H$_2$O as our case study and consider cross sections in the HST G141 bandpass (1.1-1.7 $\mu$m). Observations of exoplanetary spectra are conducted over a wide range of spectral resolution, ranging from broadband photometric observations and low resolution spectra (R $\lesssim$ 100), e.g. with HST, to very high resolution spectra with large ground based facilities (R $\sim$ 10$^5$). Resolution can greatly influence how an observed spectrum can be interpreted and can break the degeneracies between different molecules. Here we discuss the effect resolution has on $\delta$ as a function of pressure and temperature. We find that at the highest resolutions and lowest temperatures (R=10$^5$ and T=500K) pressure broadening can introduce a difference to the final cross section of $\delta$=1000\% for P=0.1 atm. For lower resolutions of R=5000, similar to those that will be attainable with JWST, the differences become much smaller. However at low temperatures (T=500K) and high pressures (P=1 atm) a $\delta$ of 40\% is found for H$_2$ broadening at R=5000. For low resolution spectra (R$\lesssim$100) of exoplanets, that are possible with current instruments, for representative exoplanetary temperatures (T=500K-2500K, P$\lesssim$1 atm) and H$_2$ rich atmospheres, we find the median difference in cross sections introduced by various aspects of pressure broadening ($\delta$) to be $\lesssim$1\%.

For illustration, Fig.~\ref{molzoom} shows molecular cross sections of several molecules binned to the spectral ranges and resolutions achievable with HST (WFC3 G102 and G141 grisms, R$\sim$100) and JWST (NIRSpec, R $\sim$3000). We can see at the higher resolution of JWST it is much easier to break the degeneracies and identify molecules. This is also easier in JWST due to the longer spectral range giving more potential to search for molecules. 

\begin{figure}
  \centering
  \includegraphics[scale=0.45]{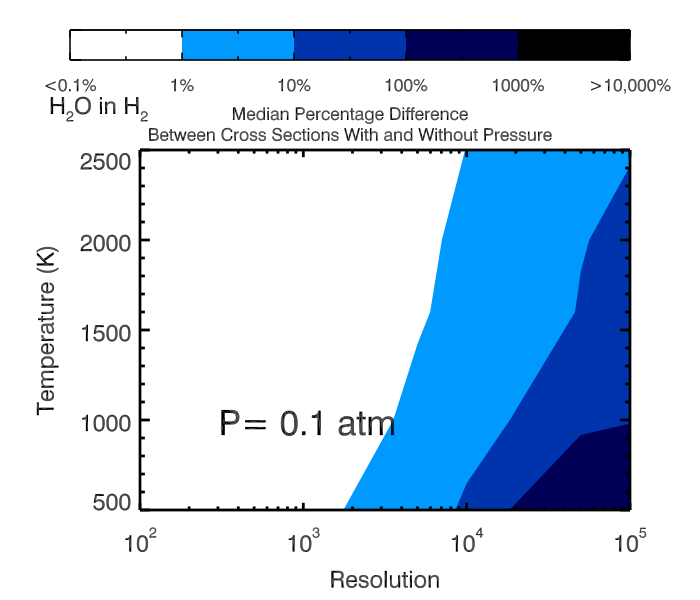}
  \includegraphics[scale=0.45]{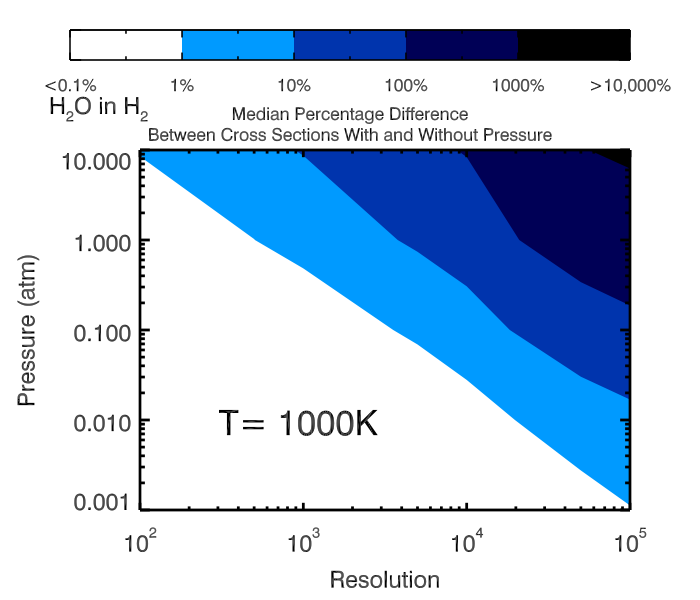}
  \caption{Effect of resolution on median difference between H$_2$O cross sections when pressure broadening is included compared with a Gaussian-only case across the WFC3 band pass range of 1-1.7 $\mu$m. Here we show H$_2$O broadened by H$_2$ though stronger broadening can be achieved using self broadening as discussed in section \ref{sec:broadening_agent}. (Note: Here the individual points of the T,P and R grid have been linearly interpolated over for plotting purposes. \citep{2013Icar..226.1673H} discusses interpolation between temperature and pressure points.)}
  \label{fig:instrumentresolution}
\end{figure}

Figure~\ref{fig:instrumentresolution} shows $\delta$ as a function of P and T. We find that $\delta$ increases with increasing resolution, reaching differences up to 100\% or higher for R $\gtrsim$ 10$^4$, P $\gtrsim 0.1$ atm, and T $\lesssim$ 500 K. For example, as shown in the upper panel of Fig.~\ref{fig:instrumentresolution} considering a temperature of 500 K, we find that at a nominal pressure of 1 atm, $\delta$ can be as high as 100\% for R ~10$^4$ or more. On the other hand, for the highest resolutions possible today of R $\sim 10^5$, $\delta \gtrsim 100\%$ even for pressures as low as 0.1 atm.

\begin{figure*}
  \centering
  \includegraphics[scale=0.2]{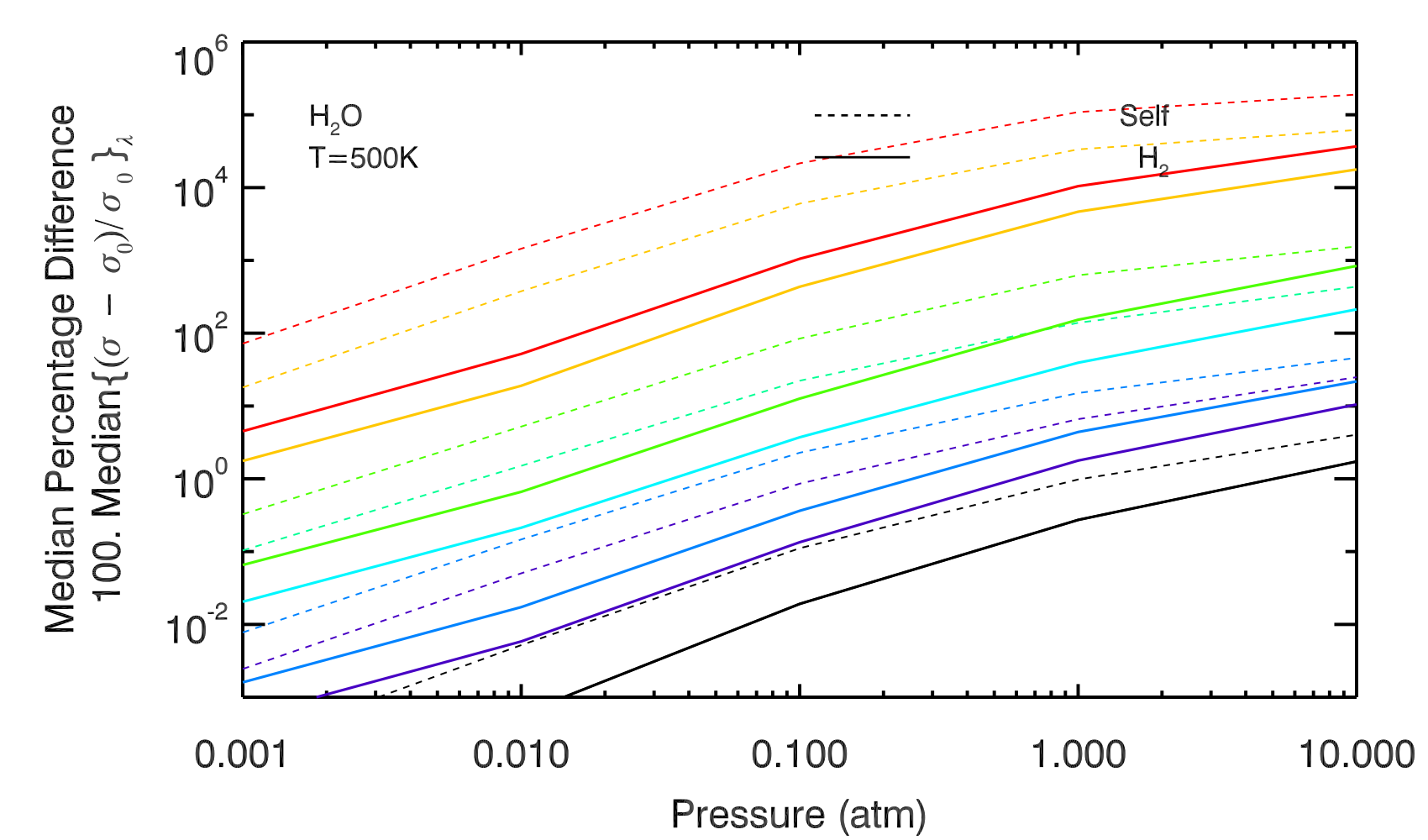}
  \includegraphics[scale=0.2]{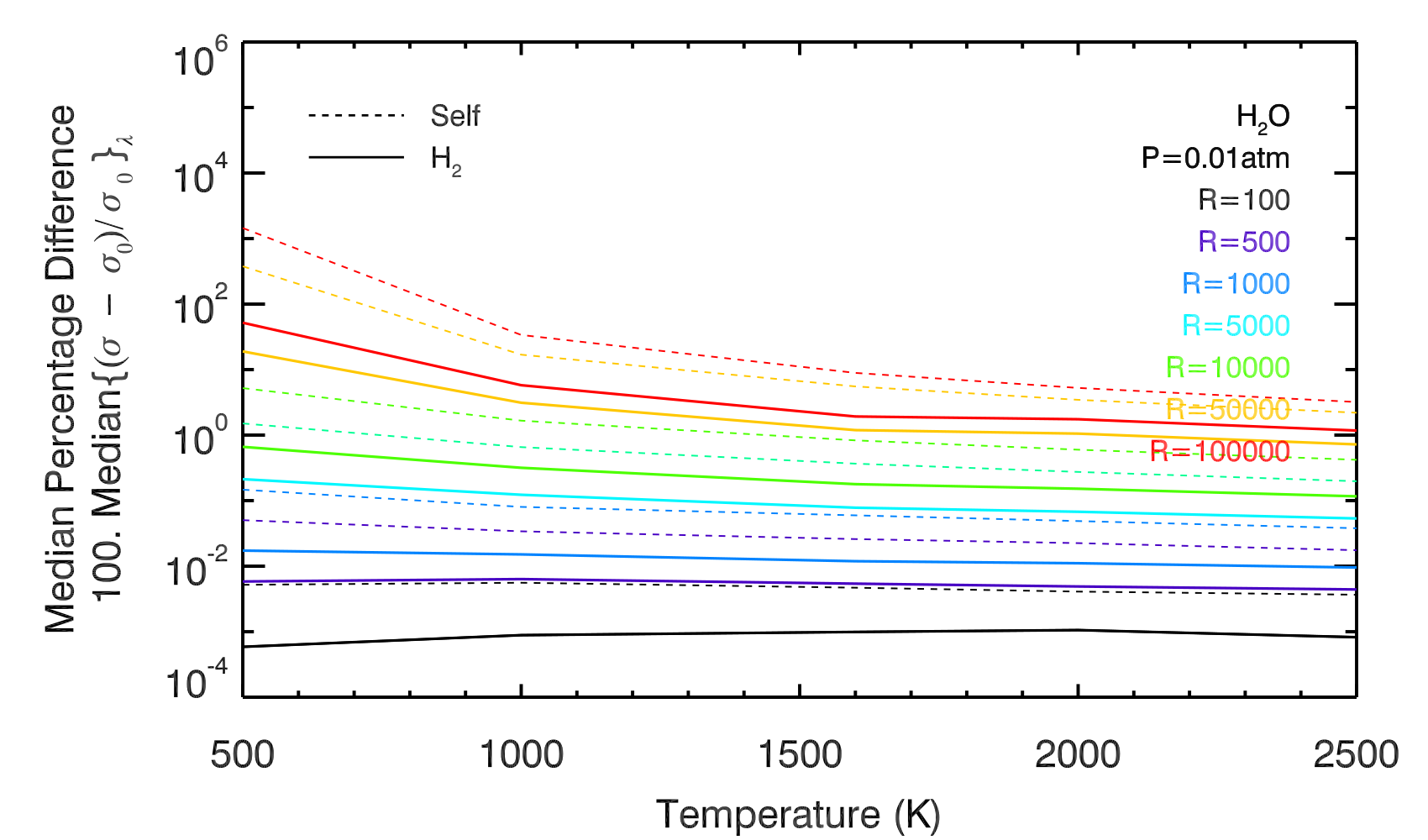}
  \includegraphics[scale=0.2]{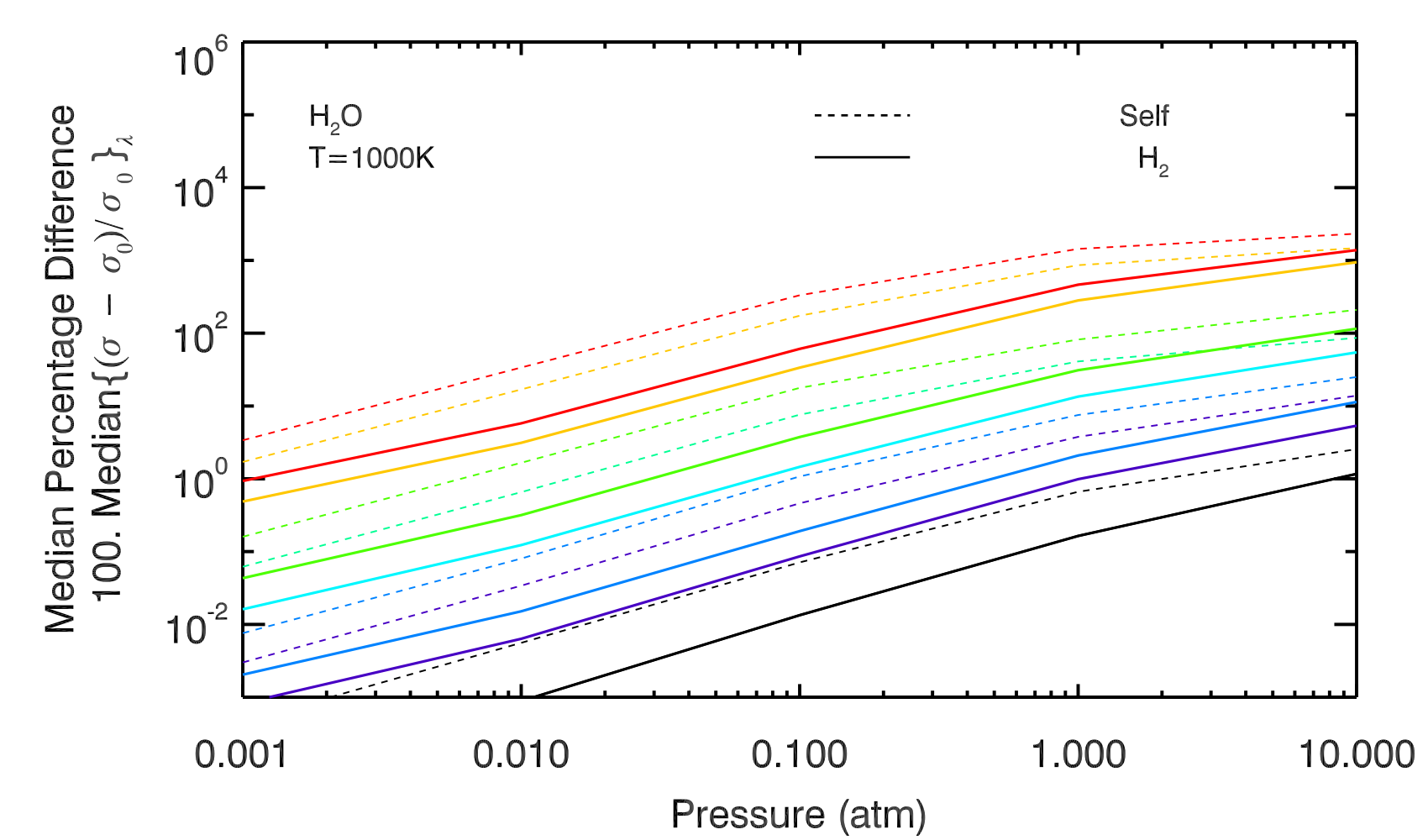}
  \includegraphics[scale=0.2]{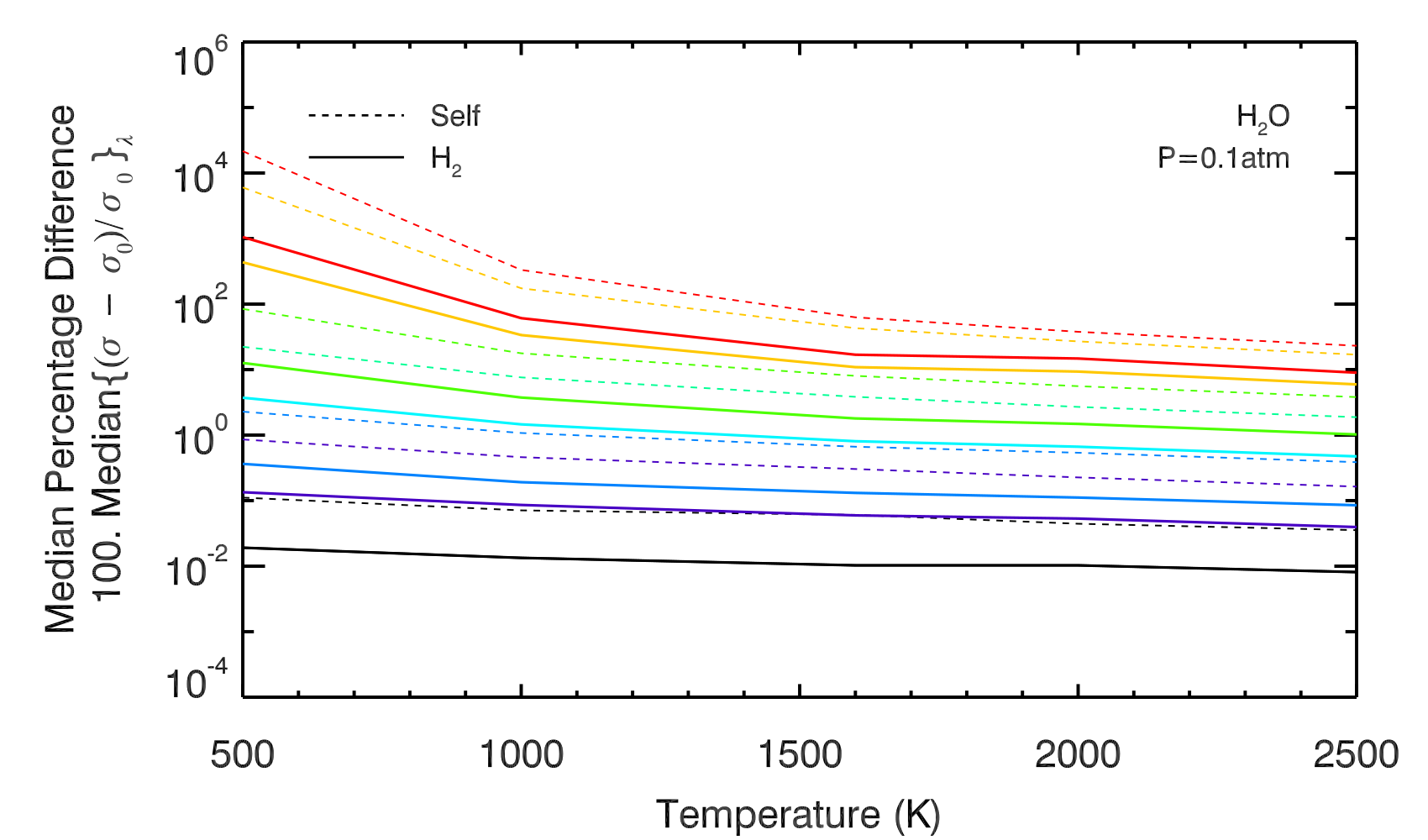}
  \includegraphics[scale=0.2]{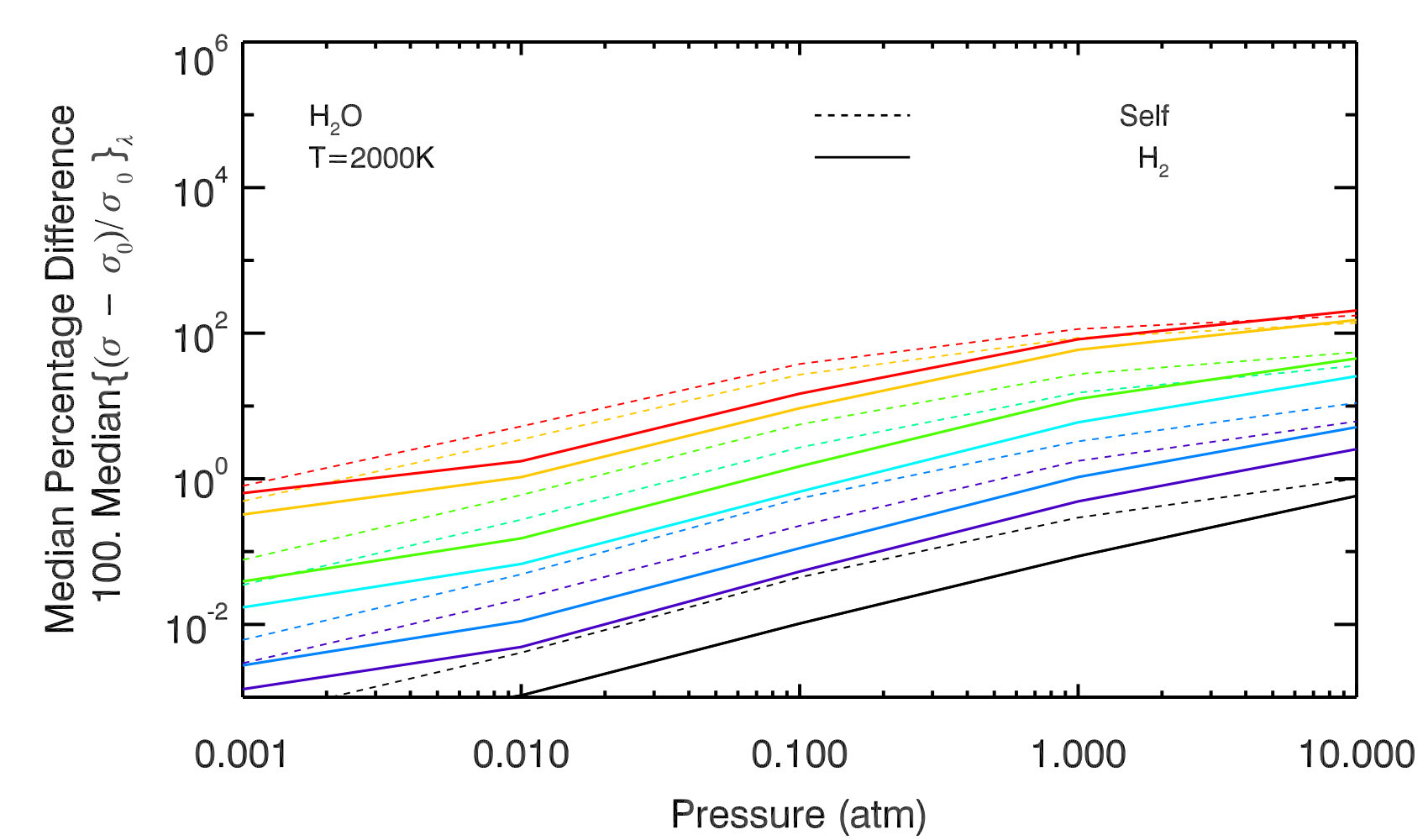}
  \includegraphics[scale=0.2]{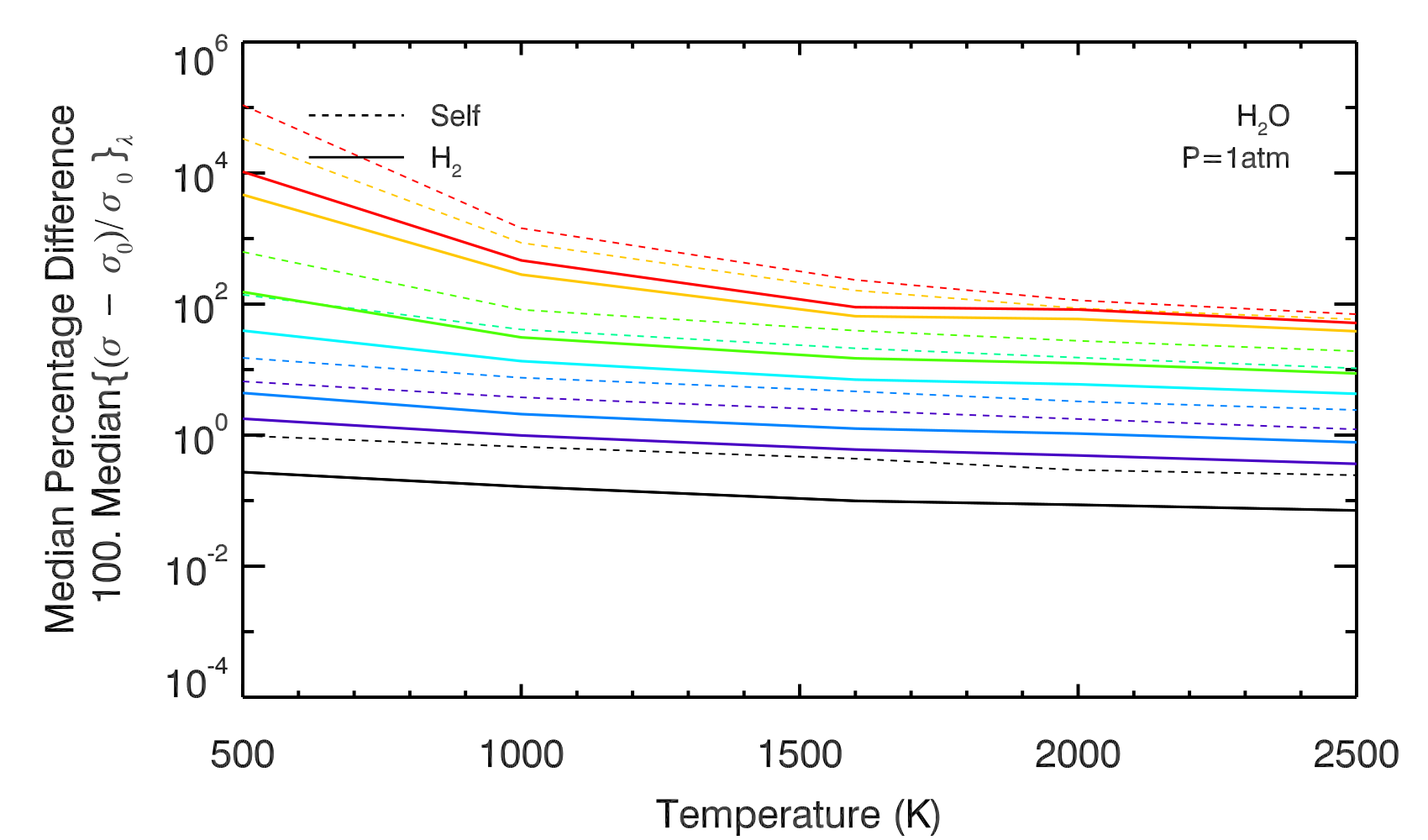}
  \caption{Effect of resolution at different temperatures and pressures on H$_2$O cross sections in the WFC3  bandpass. Median percentage difference between cross sections evaluated including pressure and cross sections evaluated with a Gaussian-only, thermal broadening. At high resolutions, high pressures and low temperatures there is the largest change to cross section. At the highest resolution of R=100,000 the median difference can be more than 1000\% for H$_2$ broadening. We see that H$_2$ broadening of H$_2$O is consistently weaker than self broadening. Resolutions of between R=100 and R=100,000 are shown in colours given by the top right panel.}
  \label{fig:instrumentresolution2}
\end{figure*}

Similarly, the lower panel of Fig.~\ref{fig:instrumentresolution} shows the variation in $\delta$ with temperature for a nominal pressure of 1 atm, showing $\delta$ can be very high ($\gtrsim 100\%$) for T $<$ 1000 K for R $> $10$^4$. Consequently, we find that it is very important for atmospheric models to include pressure broadening when interpreting high-resolution spectra (R$\gtrsim 10^4$) of exoplanet atmospheres observable with current and upcoming facilities (e.g. VLT, Keck, and E-ELT). Otherwise, the derived molecular abundances will be limited by a minimum uncertainty of more than 100\% due to inaccurate cross sections. On the other hand, we find that for R $< 10^4$, $\delta$ is reduced reaching a maximum of $\sim$ 10\% for pressures of relevance to exoplanetary atmospheric observations of P $\lesssim$ 0.1 atm. Figure~\ref{fig:instrumentresolution2} shows many slices across these plots with both H$_2$ broadening and self broadening, the latter having an even larger effect on the cross sections than H$_2$ broadening (as discussed in section~\ref{sec:broadening_agent}).  

While our results indicate that it is important to include pressure broadening in cross sections for interpreting high resolution observations, it is nevertheless advisable to also include the same for low resolution spectra as well. Even though $\delta$ is found to be at a maximum of $\sim$ 10\% for R $< 10^4$ P $\lesssim$ 0.1, low resolution spectra at very high precision with HST and JWST could allow retrieval of molecular abundances with uncertainties of a few percent, in which case the 10\% uncertainty in the cross sections could become a limiting factor. Secondly, it is to be noted again that $\delta$ is a metric of differences only in a median sense while individual lines could potentially contribute higher $\delta$ than the median value. Finally, while the current analysis focused on H$_2$O with H$_2$ broadening the same with other molecules could in principle lead to higher $\delta$ even at low resolutions which future studies need to investigate.  

\subsection{Effect of Broadening Agent}
\label{sec:broadening_agent}

Another important factor in pressure broadening is the primary broadening agent in the planetary atmosphere. As shown in Eq.\ref{eq:lorwidth}, the broadening agent governs the Lorentzian HWHM of the broadening profile. Here we find that for resolutions of R=100,000 the difference between broadening in an H$_2$ atmosphere and broadening in an $H_2$O atmosphere can be significant for water features such as those in the WFC3 1-1.7 $\mu$m bandpass. At low temperatures of 500K we find $\delta_{H_2} \sim$1000\% and $\delta_{H_2O}\sim$10,000\% for pressures of 0.1 atm at R=100,000. At higher temperatures, which have been shown to reduce the $\delta$ found in previous sections, we find that for T=1000K $\delta_{H_2} \sim$60\% and $\delta_{H_2O}\sim$300\% in the same conditions. We find that across pressure, temperature and resolution the $\delta$ from self broadening in H$_2$O is on average 4 times greater than that from $H_2$. 

Molecular line lists containing pressure broadening data, e.g. HITRAN or HITEMP, typically contain data for self and air as the broadening agents, motivated by the terrestrial applications which HITRAN was originally intended for. However, for giant planetary atmospheres H$_2$ is the dominant broadener and is of particular relevance for studying atmospheres at high spectral resolution and photometric precision. Accurate line-by-line H$_2$ broadening data for high temperatures are still elusive for most molecules of interest though a few molecules have data available, particularly for H$_2$O \citep{PS} and more recently for CO \citep{2015ApJS..216...15L}.

\begin{figure}
   \centering
   \includegraphics[scale=0.45]{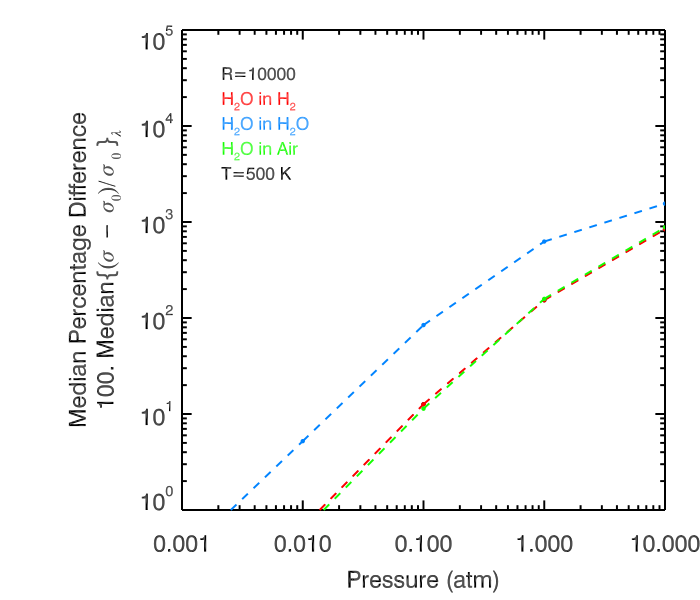}
   \includegraphics[scale=0.45]{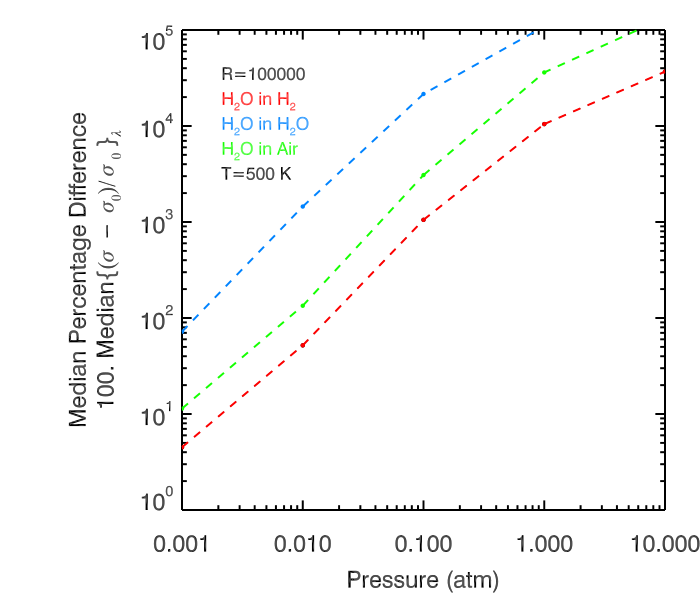}
   \caption{Effect of broadening agent on cross sections. The curves show median difference pressure broadening induces when compared with Gaussian-only cross sections when broadening agent is changed. Here the H$_2$O molecule is considered at R=10,000 and R=100,000 at a temperature of 500 K, chosen since  lower temperatures provide the highest effect from pressure broadening. The figure illustrates that self  broadening is 400\% stronger on average than $H_2$ broadening and that air and $H_2$ broadening are comparable in magnitude.}
   \label{molfracdiff}
 \end{figure}

Here, we investigate the effect of broadening agent on the median accuracy of molecular cross sections for a representative case. We consider the case of H$_2$O for which we have line-by-line broadening parameters with H$_2$, self, and air as broadening agents \citep[][R. Freedman - personal communication]{PS}. To illustrate the differences made by changing the broadening agent we have used cases where the molecule is broadened only by a particular molecule self, air, or H$_2$ (i.e. the partial pressure is 1 in each case). Figure~\ref{molfracdiff} shows the median percentage difference in cross sections caused by each of the three scenarios compared to Gaussian-only broadening for an illustrative case with resolutions of 10$^4$ and 10$^5$ and T = 500 K.

Figure~\ref{molfracdiff} shows that it is important to carefully choose broadening agents before generating cross sections for different planet types. Firstly, self-broadening can cause significantly higher $\delta$ values compared to air or H$_2$ broadening at observable pressures (P $\sim$ 0.1 - 1 atm). Secondly, the differences between H$_2$ broadening and air broadening are relatively small in the H$_2$O case. Therefore, while modelling H$_2$-rich atmospheres in the absence of any H$_2$ broadening data for H$_2$O molecules, though not ideal, it is more advisable to use air broadening than self broadening. On the contrary, when modelling atmospheres of low-mass exoplanets, e.g. super-Earths that can have volatile-rich atmospheres such as H$_2$O-rich or CO$_2$-rich atmospheres, it is important to use cross sections that are generated with the appropriate broadener. For example, for H$_2$O-rich atmospheres self-broadening of H$_2$O should be considered in the cross sections rather than air broadening. 

The effect of broadening agent can be substantial depending on other parameters. The effect of the broadening agent is naturally strongest in regions of parameter space where pressure broadening is expected to be strongest, namely at high pressures, low temperatures, and high resolution. Fig.~\ref{fig:instrumentresolution2} shows the differences between self and H$_2$ broadening for various parameters. The differences begin to become significant at high pressures (P $\gtrsim$ 0.1 atm) for R $\gtrsim$ 10000 and become substantial even at lower pressures for high resolutions. For very high resolution observations at R $\gtrsim$ 10$^5$ self broadening can lead to differences of 1000\% in cross sections. This has a much greater effect than H$_2$ broadening over a large range of pressures, making it critical to use broadening data with the appropriate broadening agents for interpreting observations at these resolutions. Finally, the differences in cross sections induced by different broadening agents are particularly strong at lower temperatures (T $\lesssim$ 1000 K) as the broadening has a greater effect at these temperatures. 

Given the wide range of temperatures and pressures of exoplanetary atmospheres that are accessible to current and upcoming observations line-by-line broadening parameters are required in molecular line lists for different broadening agents. In particular, there is a critical need for high-temperature ($\gtrsim$ 500 K) H$_2$ broadening data as the most observable atmospheres are those of hot and giant exoplanets with H$_2$-rich atmospheres, for which very high resolution spectra (R $\sim$ 10$^5$) are also being reported. In the mean time, it is advisable to use air broadening where available for such atmospheres because while not ideal it provides closer cross section estimates to H$_2$ broadening for H$_2$O and are an improvement on incorporating no pressure broadening at all. Future studies would also need to investigate if the same is true for other molecules as and when H$_2$ broadening data become available for those molecules. Finally, a realistic atmosphere will contain many different molecules and so contribute broadening from many different species. For smaller planets we expect atmospheres that are more complex, containing more massive molecules with high abundance. This will effect the pressure broadening particularly as the partial pressure will no longer be 1 and there will be contributions from many species. In such cases we expect the contribution to vary depending on the abundance of more massive broadening molecules with greater pressure broadening parameters. Our current work gives an estimate for the most extreme cases of H$_2$O or H$_2$ dominated atmospheres. 

 \begin{figure}
   \centering
   \includegraphics[scale=0.45]{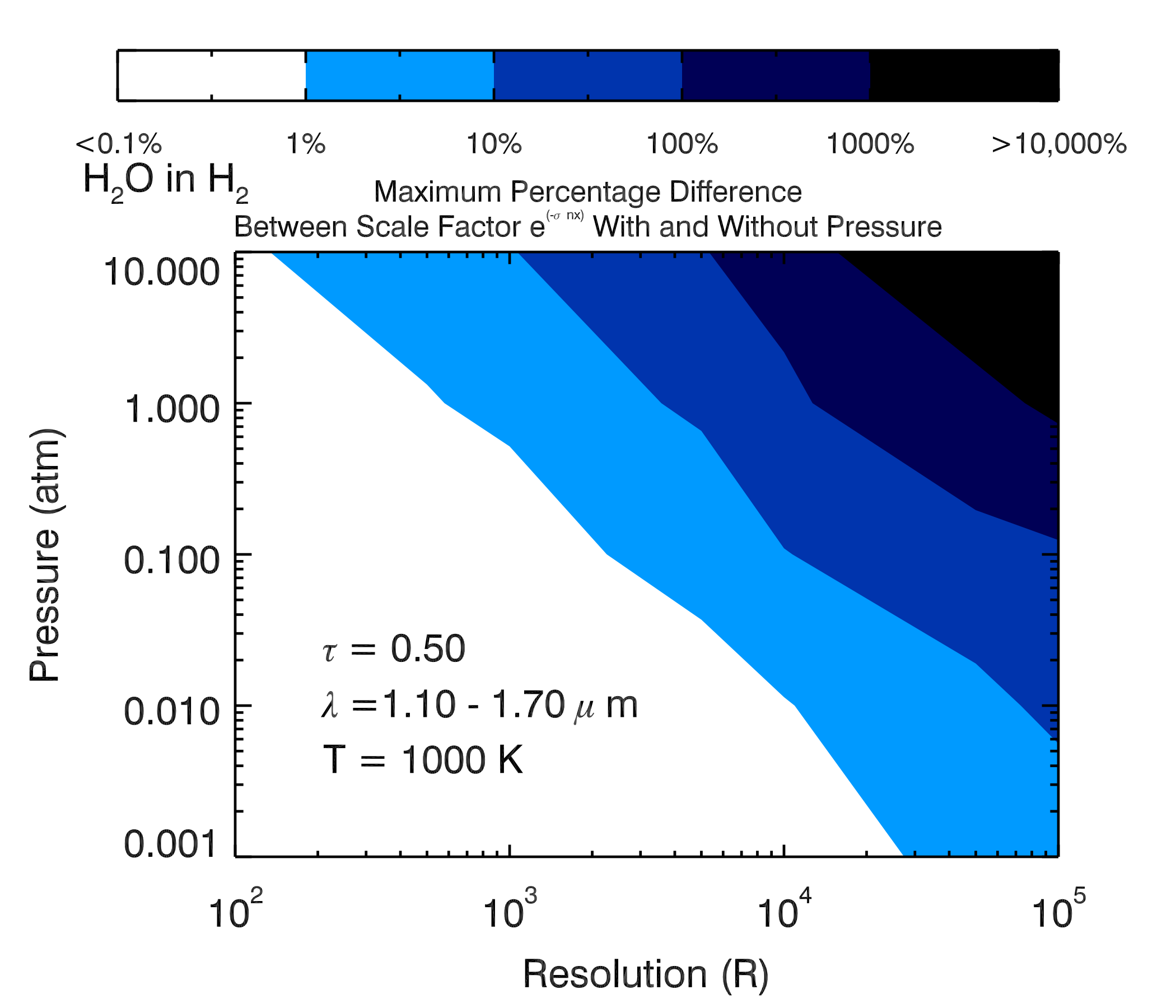}
   \includegraphics[scale=0.45]{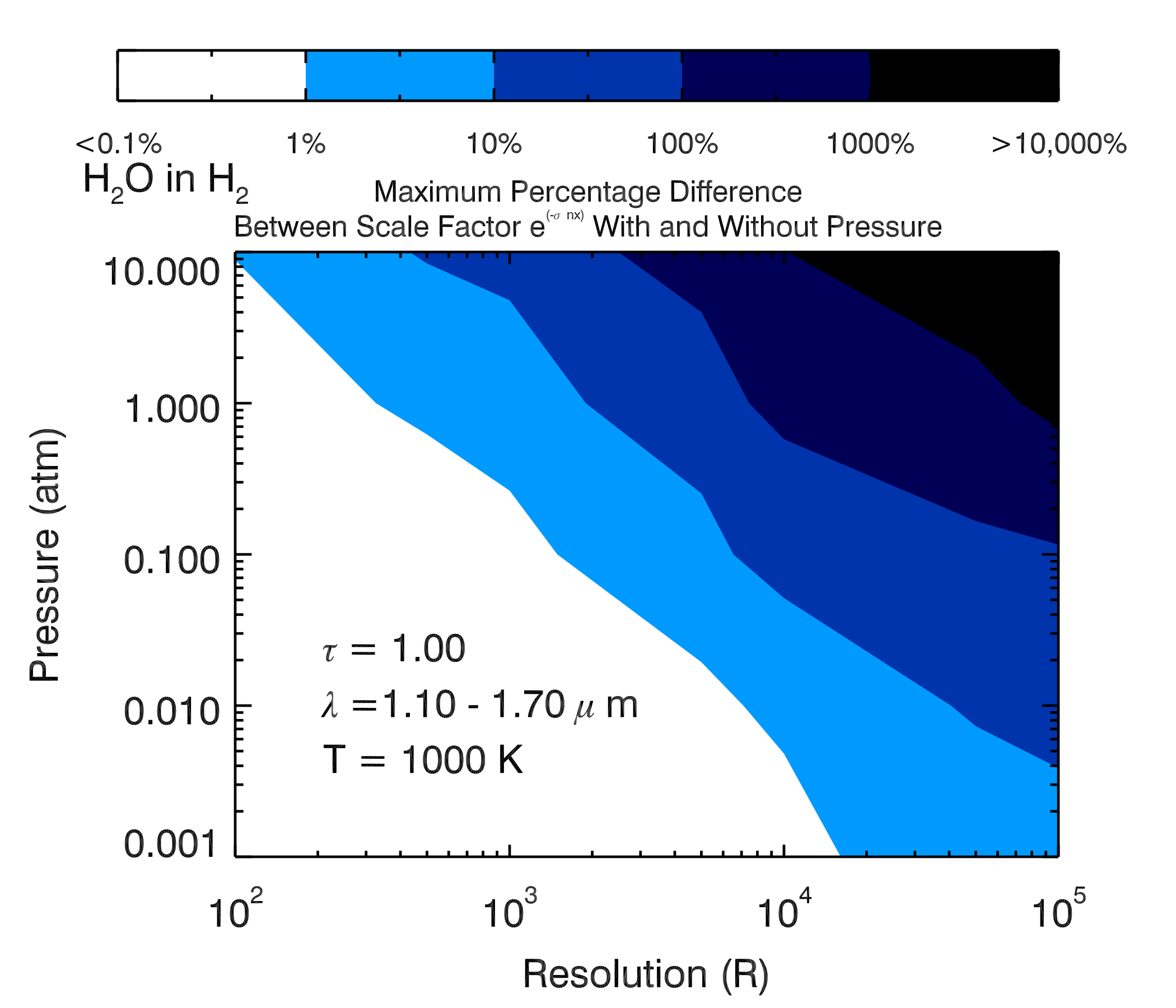}
   \includegraphics[scale=0.45]{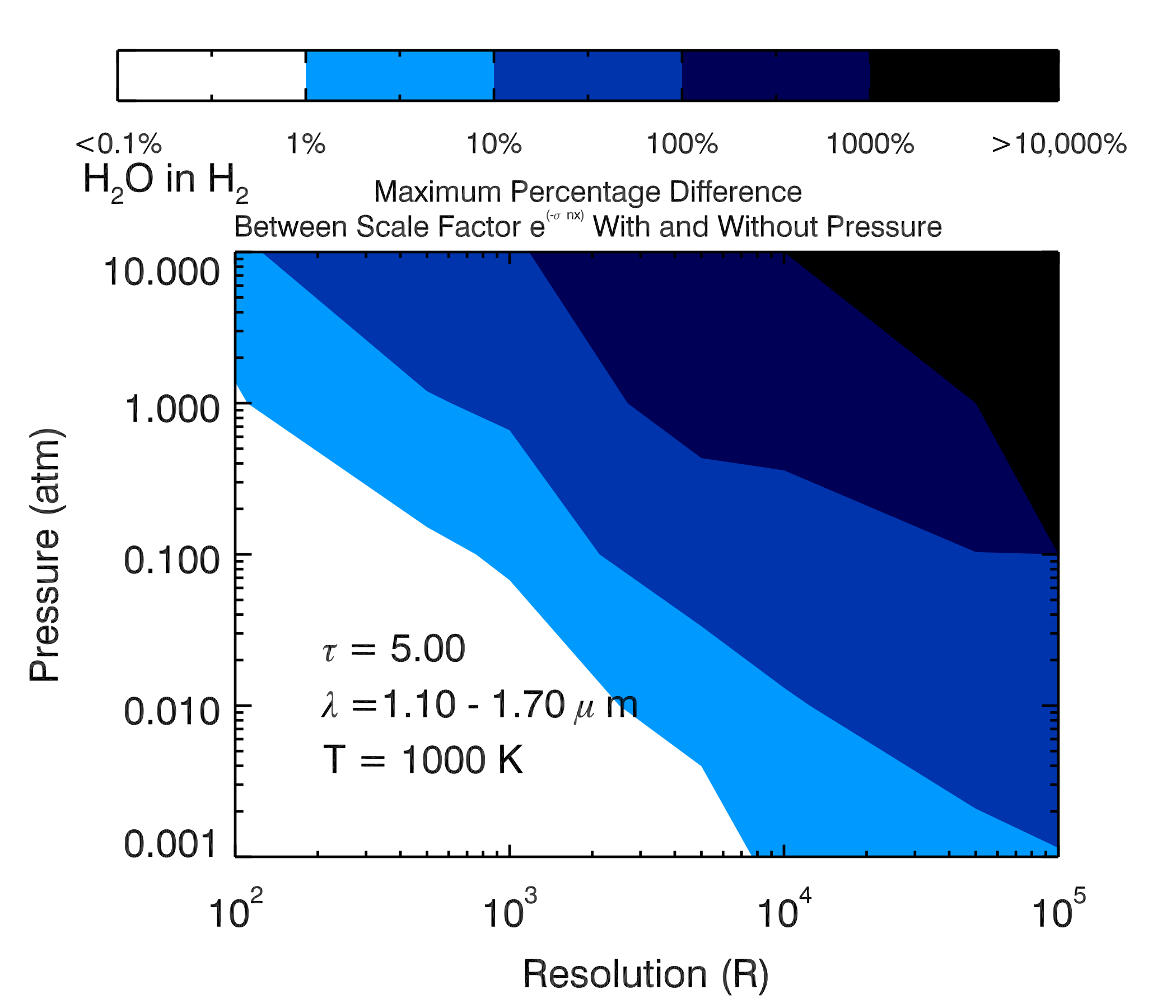}
   \caption{Effect of pressure broadening on the transmittance in an idealized atmosphere (see section~\ref{sec:optical_depth}). The contours show the maximum percentage difference in transmittance, i.e. the scale factor $e^{(-n \sigma x)}$, induced by considering cross sections with pressure broadening versus those with Gaussian-only thermal broadening in the HST WFC3 bandpass. The relative difference is shown for a wide range in key parameters: pressure (P), spectral resolution (R), and maximum optical depth in the WFC3 bandpass ($\tau$); a nominal temperature of 1000 K is chosen for illustration but the general temperature-dependence is discussed in section~\ref{sec:optical_depth}. Here the individual points of the T,P and R grid have been linearly interpolated over for plotting purposes \citep[see e.g.][]{2013Icar..226.1673H}} 
   \label{tau}
 \end{figure}

\section{Effect of Pressure Broadening on Transmittance}
\label{sec:optical_depth}

A thorough investigation of the effect of pressure broadening on fully modelled exoplanetary spectra is non-trivial and beyond the scope of the present study. Many factors such as the inhomogeneous P-T structure and composition of the atmosphere will determine the final spectrum. Nevertheless, as a simpler exercise, here we nominally asses the effect of pressure broadening on the transmittance in a fiducial atmosphere represented by a uniform column of gas. As in other parts of this study we use H$_2$O as the only absorbing species, and we consider pressure broadening in a H$_2$-rich atmosphere.

Cross sections are used in atmospheric codes to determine the resultant intensity transmitted through a column of gas. Usually, the  column will undergo some changes in pressure, temperature, and number density over the length of the column. These factors combine to give an optical depth $\tau_\lambda$ where

\begin{equation}
  \tau_\lambda=\int_{z_1}^{z_2} n\sigma_\lambda dz
\end{equation}

where z is length through the column, n is the abundance (number density) of the molecule and $\sigma_\lambda$ is the cross section which is a function of wavelength. The abundance of a molecule is determined by the temperature and pressure of the column by the simple gas law
\begin{equation}
  P={nk_BT}.
\end{equation}

 The optical depth gives a measure of how much intensity will be transmitted through a column of gas based on these properties. For a source intensity ($I_{0\lambda}$), the resultant intensity ($I_\lambda$) at the end of the column is given by

\begin{equation}
  I_\lambda=I_{0\lambda}e^{-\tau_\lambda}, 
\end{equation}
where, the scale factor $e^{-\tau_\lambda}$ is the transmittance.

 In order to make a meaningful comparison between the effects of altering the cross section on emergent intensity as a function of pressure and temperature we choose to fix $\tau$ to a single value. Here, we assume the column of gas to be at a given constant pressure and temperature, and hence constant density. The length of the column is allowed to vary in order to contain the same $\tau$ regardless of pressure and temperature. As $\tau$ is also a function of wavelength its value alters depending on the particular molecular feature. To fix $\tau$ we take the value at the peak of the water feature near 1.4 $\mu$m in the WFC3 band. If only the cross section $\sigma_\lambda$ is changed, between a pressure broadened case $\sigma_{1\lambda}$ and an unbroadened case $\sigma_{2\lambda}$ which will each have intensities $I_{1\lambda}$ and $I_{2\lambda}$, we can then find the effect pressure broadening has on the transmitted intensity. This is given as 

\begin{equation}
  \frac{I_{1\lambda}-I_{2\lambda}}{I_{2\lambda}}=\frac{e^{-\int_{z_1}^{z_2} n\sigma_{1\lambda} dz}-e^{-\int_{z_1}^{z_2} n\sigma_{2\lambda} dz}}{e^{-\int_{z_1}^{z_2} n\sigma_{2\lambda} dz}}
\end{equation}
As we assume a gas of the same number density at each pressure and temperature point then we can assume
\begin{equation}
  A=\int_{z_1}^{z_2}n dz=nx
\end{equation}
where $x$ is some distance scale. Substituting this we find
\begin{equation}
  \Delta I_{\lambda}=\frac{I_{1\lambda}-I_{2\lambda}}{I_{2\lambda}}={(e^{-A(\sigma_{1\lambda}-\sigma_{2\lambda})})-1}.
\end{equation}

The above expression gives the relative change to the intensity, and hence the transmittance, for a given abundance and path through a uniform column of gas induced by using cross sections with pressure broadening compared to those without. Using the cross section across the WFC3 bandpass of 1-1.7$\mu$m and binning down to a given resolution we can find the difference to the transmittance of the column of gas as a function of wavelength. As discussed above, the length of the column is fixed such that the maximum optical depth of the column in the given bandpass equals a fixed parameter ($\tau$), for a given density corresponding to a given temperature and pressure. We can then alter $\tau$ to investigate the optically thin and optically thick regimes as functions of pressure and temperature.  We note that the change induced to transmittance ($\Delta I_{\lambda}$) across a given bandpass is higher at wavelengths with higher absorption, which are also of the wavelengths of interest to observations. We therefore consider the max($\Delta I_{\lambda}$) in the WFC3 bandpass as our metric of choice in evaluating the effect of pressure broadening on transmittance in that bandpass. This does not take into account how signal to noise might affect taking such observations as zero transmittance ($e^{-\tau}$) implies no signal;  however, here we consider only values of $e^{-\tau}$ (which can have values between 0 and 1) that are greater than 0.01. 

 The effect of pressure broadening on the transmittance in our idealized column of gas is similar to the effect on cross sections discussed in previous sections. Figure~\ref{tau}  shows the fractional difference ($\Delta I_{\lambda}$) pressure broadening makes to the transmittance as a function of several key parameters: the optical depth ($\tau$), pressure (P), and resolution (R); a nominal temperature of 1000 K is chosen for illustration but the general temperature-dependence is discussed below. At the outset, for low resolutions (R$\lesssim$100), $\Delta I_{\lambda}$ is $\lesssim$1\% across the HST WFC3 G141 bandpass (1.1-1.7 $\mu$m) for almost the entire range of parameters of relevance to exoplanetary atmospheres, particularly for P$<$ 1 atm, T = 500 - 3000 K, and $\tau < 5$. 
Naturally, however, $\Delta I_{\lambda}$ is higher for higher resolutions. Considering nominal values of $\tau<1$, P$<$0.1 atm, and T$>$1000K, we find a maximum $\Delta I_{\lambda}$ in the WFC3 band to be 6\% for a JWST-like medium resolution of R = 5000, and 75\% for a VLT-like very high resolution of R = $10^5$. 

The $\Delta I_{\lambda}$ for each resolution increases with increasing pressure and lowering temperature, particularly for very high resolution. For our lowest T of 500 K and $\tau=1$, for R = 5000 $\Delta I_{\lambda}$ is $\lesssim$12\% for P$<0.1$ atm, and $\lesssim$  65\% for P$<1$ atm. On the other hand, for the same T and $\tau$, for R = $10^5$ $\Delta I_{\lambda}$ is $\lesssim$ 100\% for P$<$ 0.1 atm and $\lesssim$  2000\% for P$<1$ atm. As $\tau$ increases the difference between the two cases increases as there is more material to modify the intensity. However, as $\tau$ is increased to very high values the medium becomes optically thick and no light is transmitted in certain wavelength regions. 

This approach is simplistic as clearly it does not factor in the the changes that could happen within the column in temperature and pressure however this does give us a first approximation of the difference induced by changing cross sections on observations of transmission spectra of exoplanetary atmospheres. In reality, light travels through many layers of an exoplanetary atmosphere, with different temperatures, pressures, and densities, before reaching the observer. The results above will hold for a specific pressure but full spectral models of exoplanetary atmospheres, both for transmission spectra and emission spectra, are required for a comprehensive investigation of the effect of pressure broadening discussed in the present work.

\section{Cross-Section Database}

In this work we present a range of cross sections for H$_2$O, CO$_2$, CO, CH$_4$, NH$_3$ and HCN from a range of sources shown in Table~\ref{tab:allmols}. These have been investigated with detailed, line-by-line calculations of the Voigt profile with pressure and thermal broadening simultaneously included which no other database to date provides. The cross sections span a temperature range of 300K-3500K and pressures of 10$^{-4}$ to 10$^2$ atmospheres. Finally the cross sections have been created in a variety of resolutions.

The data in this work benefits not only from the addition of a further dimension of pressure with an accurate broadening profile, but also in being generated uniformly with the same code across molecules. This ensures low and consistent systematic errors across our data. The full database is represented in Figure~\ref{fig:AllMolecules}. 

\subsection{Molecular Spectra of Observational Relevance}
\begin{figure*}
  \centering
  \includegraphics[scale=0.9]{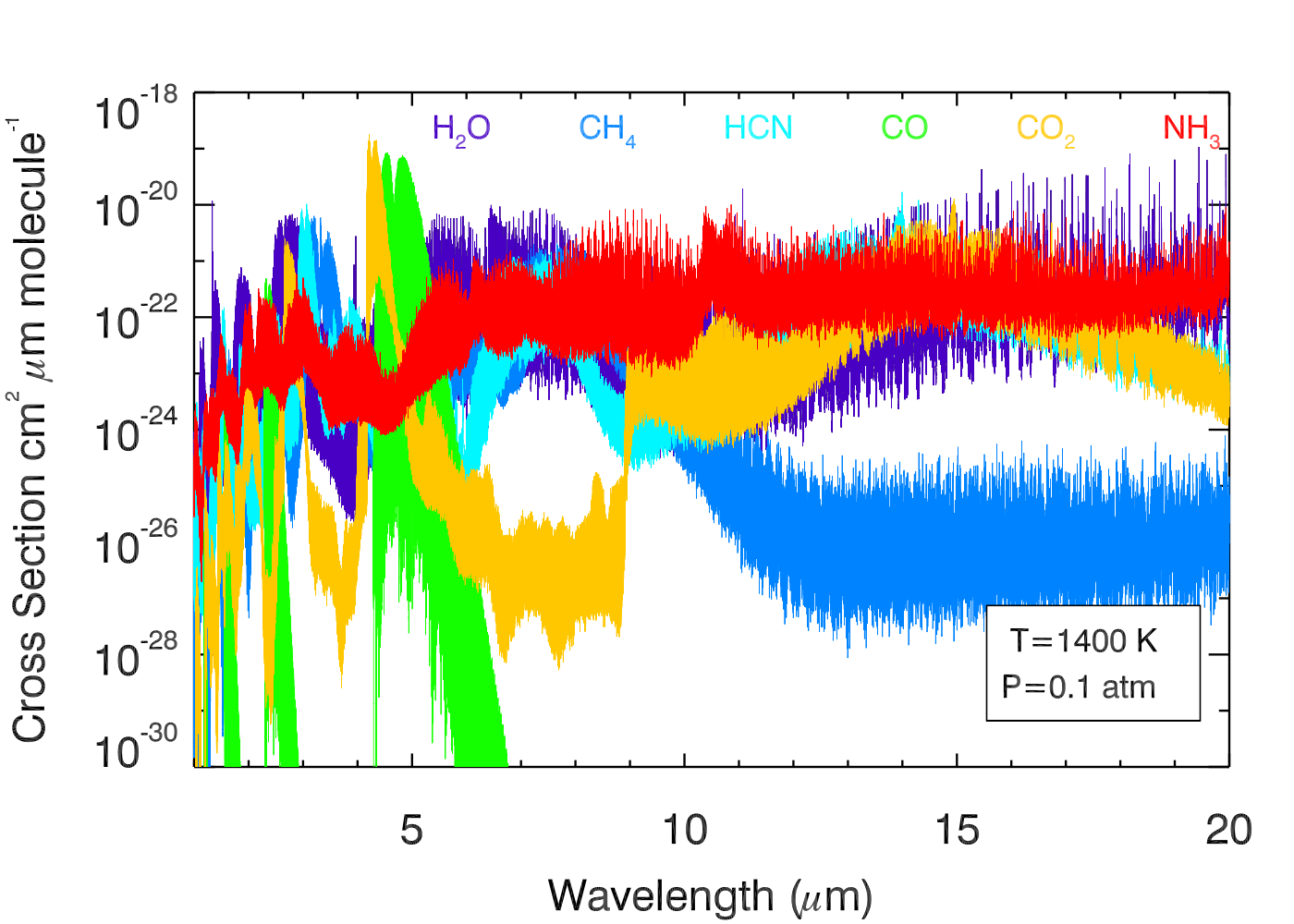}
  \caption{Absorption cross sections of molecules in our database for a representative T=1400 and P=0.1 at a resolution of 0.01 cm$^{-1}$ wavenumbers, using air broadening, with the same relative abundances. Such cross sections have been generated for all the molecules over a wide range of P and T using different sources of line lists and broadening molecules. Sources are listed in Table \ref{tab:allmols}.}
  \label{fig:AllMolecules}
\end{figure*}

Figure~\ref{fig:AllMolecules} shows cross sections for each molecule that has currently been addressed in this work. Many of these molecules have had cross sections computed for different line lists, broadening molecules and for a mean or detailed approach but here only the most complete line list cases are shown. 

Each of the cross sections contains strong molecular features relating to the particular molecule. These features are usually the most ideal regions for observations with low signal to noise. A list of the highest intensity features is given in Table~\ref{FeaturePositions} with their representative central wavelengths.

\begin{center}
  \begin{table}
    \centering
    \caption{Positions of the most prominent features in the absorption cross sections of each molecule.}
    \begin{tabular}{cl}
      \hline
      \textbf{Molecule} & \textbf{Feature Position $\mu$m} \\\hline
      H$_2$O	&	6.61, 5.90, 2.76, 2.67, 2.60 \\
      & 1.87, 1.36, 1.13, 0.95	\\\hline
      CO$_2$	&	14.95, 4.23	\\\hline
      CO	&	4.57, 4.32, 2.68, 2.00	\\\hline
      HCN	&	14.00, 7.30, 6.93, 4.73, 3.86 \\
      & 3.57, 3.00, 2.50, 1.53	\\\hline
      NH$_3$	&	15.96, 12.14, 10.37, 9.23, 6.67 \\
      & 6.15, 3.00, 2.26, 1.95, 1.51 \\
      & 1.22, 1.03	\\\hline
      CH$_4$ & 7.70, 7.40, 6.47, 3.42, 3.32 \\
      &3.21, 2.37, 1.67 \\\hline
      NO &  5.32, 2.58 \\\hline
      OH &  4.20, 2.25, 1.55 \\\hline
    \end{tabular}
    \label{FeaturePositions}

  \end{table}
\end{center}

\subsection{Temperature Dependence of Cross-Sections}

\begin{figure}
  \centering
  \includegraphics[scale=0.65]{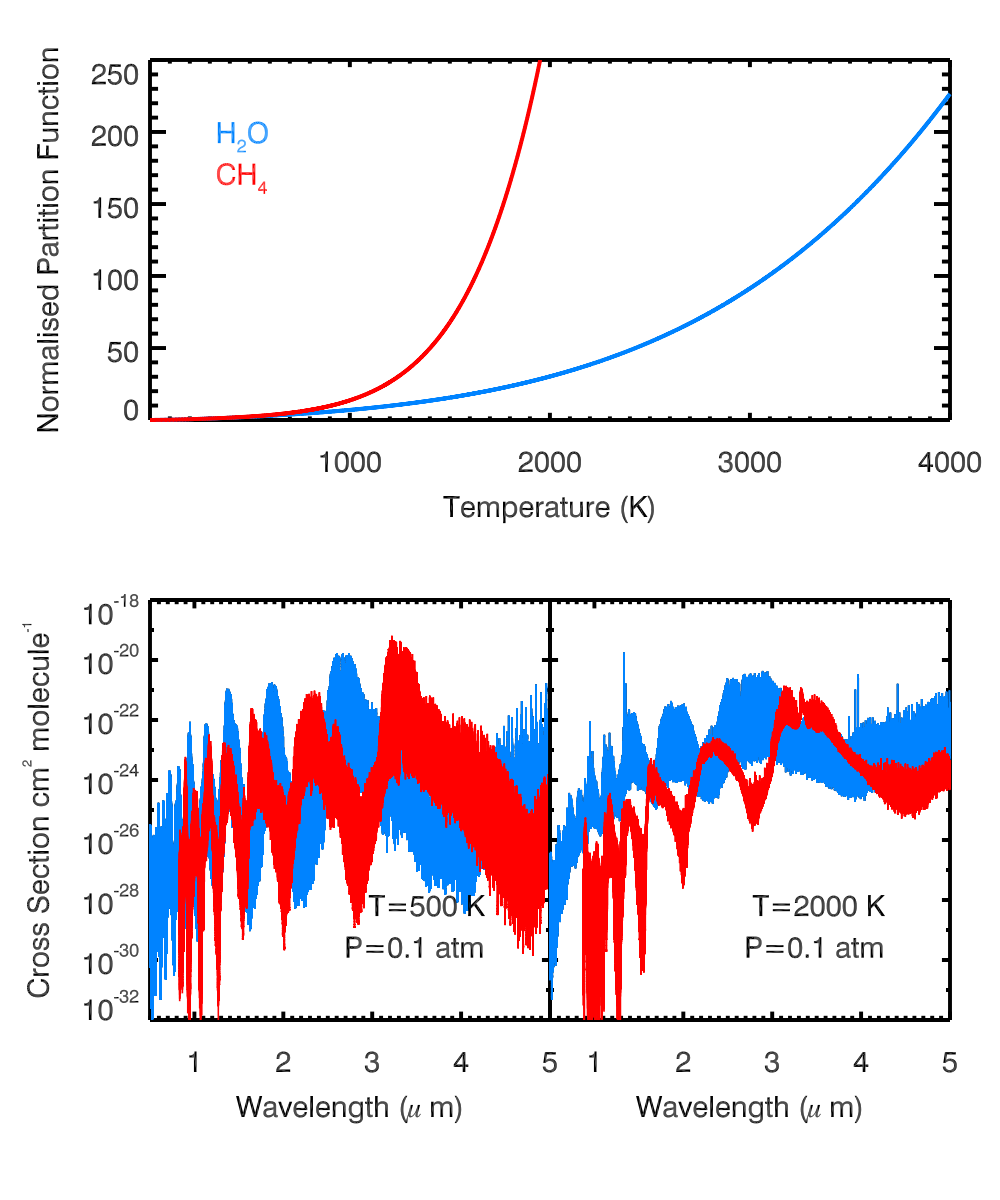}
  \caption{Comparison of CH$_4$ and H$_2$O across different temperatures. \emph{Top:} comparison of normalised partition functions ($Q(T)/Q(T=296K)$). CH$_4$ shows a much stronger temperature dependence than H$_2$O. \emph{Bottom:} Comparison of cross sections at different temperatures. We see that the contrast between peaks and troughs in both spectra is lowered at high temperatures due to the partition function. However CH$_4$ is is more effected due to the higher partition function values and, at higher temperatures, is weakened compare to H$_2$O.}
  \label{fig:CH4H2O}
\end{figure}

There is a strong dependence on temperature in these cross sections where increasing the temperature raises the transition features in the spectrum, reducing the contrast between the peaks and troughs of each band of transitions. Molecules have a different temperature dependence based on their partition function. Individual molecules can have a stronger or weaker temperature dependence which can have an affect on our observations as in an atmosphere many molecular features will be combined. An example of the effect of such a case is shown in Figure~\ref{fig:CH4H2O}. CH$_4$ has a stronger temperature dependence than H$_2$O. This leads to changes in their relative contributions to the combined cross section as temperature increases. While at low temperatures the two contributions are comparable, at very high temperatures the H$_2$O cross section is stronger than the CH$_4$ cross section. While it would still be possible to tell these two contributions apart in high resolution spectra it would be more difficult to assess the abundance of CH$_4$ in the presence of H$_2$O. This is a general effect from the partition functions, but there are subtleties depending on the individual transitions. For example, transitions with high values of the total angular momentum quantum number J, and high lower energy levels,  become relatively much stronger at higher temperatures. Such effects can influence the shape of individual bands and could affect the continuum level of low intensity `window' regions in spectra where opacity is low.

This is a demonstration of the effect and is not necessarily a likely combination of molecules, however cross sections such as those developed here give us a tool to understand at what temperatures molecules are dominant, making them easier to observe. Understanding these features and their behaviour both with wavelength and temperature is crucial when developing ideal band passes for observation, particularly when taking telluric lines into account. 

For simple molecules such as CO there are only a few key bands in the infra-red which will contain spectroscopic information while for more complex molecules there is a contribution across the whole wavelength range. The most prominent features for each molecule are listed in Table~\ref{FeaturePositions}. The WFC3 G141 bandpass (1.1-1.7 $\mu$m) is very well placed to detect and characterize H$_2$O, CH$_4$, and NH$_3$ absorption features in spectra of exoplanets and brown dwarfs at low resolution ($R \lesssim$ 100). On the other hand, the NIRSpec and MIRI instruments aboard the JWST spanning a wide range of 0.6-28 $\mu$m will be able to detect a wide range of molecules at higher resolution.

\section{Discussion and Summary}

In this work we present a systematic and quantitative investigation of the effects of various aspects of pressure broadening on molecular cross sections for application to exoplanetary atmospheres. We first use H$_2$O as our primary molecule of choice for this investigation as it has the most complete absorption line data. The factors we investigate include the resolution and evaluation width of Voigt profiles, pressure versus thermal broadening, broadening agent, spectral resolution, and completeness of broadening parameters. We investigate in detail the effect of pressure broadening both on the absorption cross sections of H$_2$O under varied conditions as well as on the transmittance of a fiducial idealized atmosphere. We use the optimal methods resulting from this investigation to systematically and homogeneously generate a library of pressure-broadened absorption cross sections for a wide selection of molecules of relevance to exoplanetary atmospheres across a wide range of temperature, pressure, and spectral resolution.

This study allows us to address the question of the inaccuracies introduced to molecular absorption cross sections from pressure broadening, both in the context of current and future observational capabilities. As new instruments come online with improved specifications we will have access to a wealth of high resolution data on exoplanet atmospheres. The interpretations of these data sets will be impacted by basic model inputs such as cross sections which are directly degenerate with the molecular abundances derived using spectral retrieval methods. The comparisons presented here show in detail the magnitude of the errors we can expect in these fundamental inputs to atmospheric models across a range of parameters. 


To generate cross sections we follow a prescription of mapping line intensities broadened by an appropriate function to a fine 'sub-grid' which finely samples the profile of the line. This is iterated over each transition from the source and binned to an output grid with a wider spacing, (lower resolution,) for further use. The lines are broadened by either a Gaussian-only model, (which uses only thermal broadening,) or a Voigt profile, (which combines the Gaussian thermal and Lorentzian pressure broadening,) evaluated using the Faddeeva package \citep{Faddeeva}.


When evaluating the Voigt profile on a grid there are two clear sources of error. Firstly the grid spacing may be too wide, causing the evaluation of the contribution to each grid point to be poor, leading also to a poor normalisation and misrepresenting the line transition intensity. Secondly the wings of the profile may be cut off prematurely, leading to small fractions of intensity from the wings being missed. This is aggravated by the range of intensity values which span many orders of magnitude within a narrow wavenumber range. This leads to the wings of isolated, high intensity profiles affecting the continuum level of low intensity neighbours greatly, which can be underestimated when a cut off is too narrow.

As discussed in section~\ref{sec:optimal_grid} we present a method of accurately evaluating the Voigt profile on a fine grid that produces minimal errors in the final cross section at resolutions of interest. This is achieved with a spacing that is adaptive in temperature and in pressure. The grid we adopt is found to be as accurate as the grid from \citep{2013Icar..226.1673H} to within ~0.2\% at pressures of 1 atm or less and gives a vast saving on computational time, particularly for high pressures. We pair this grid with a cut off value, $\Delta\nu_c$, of 500 Voigt widths (raised to 1000 above pressures of 1 atm). $\Delta\nu_c$ describes the separation around the wavenumber centroid of the line transition up to which the Voigt profile is evaluated to. Having investigated a range of values at different resolutions we find 500 Voigt widths to be sufficient both to provide good normalisation for the profile and to evaluate far into the extensive Lorentzian wings. When compared with other values of $\Delta\nu_c$ from literature we find our value to be more accurate (~10-100\%) at low pressures (P= 0.1 - 0.001 atm) due to the Voigt width adapting with both temperature and pressure as the Gaussian and Lorentzian components change. This combination of $\Delta\nu_c$ and an adaptive grid provides low errors for all resolutions discussed in this work up to R=100,000. We find errors of less than 1\% (averaging $\sim$0.2\%) in the final cross sections at the peaks of transition features. This increases to 10\% within transition features at very low intensity, however we find such transitions to be less likely to significantly effect observations and modelling results. Beyond this resolution it may be wise to increase $\Delta\nu_c$ and use a finer sampling of the profile.

Here we use a standard Voigt profile though works such as \citet{2012RSPTA.370.2495N} have shown that a change in velocity of the broadening particle can alter the pressure broadened profile shape. This would likely change the wing shape and alter the continuum from what we present here though we anticipate that difference to be small.

We find the effect of pressure broadening to be varied depending on the resolution, pressure and temperature. We choose to measure the change induced to H$_2$O cross sections due to pressure broadening using the median percentage difference over the HST WFC3 bandpass (1.1-1.7 $\mu$m). This provides a reasonable estimate of  the characteristic difference, though it is possible to induce higher changes for specific lines. Generally, the differences are larger for higher resolutions, higher pressures, and lower temperatures. For low resolution spectra (R$\lesssim$100) of exoplanets, that are possible with current instruments, for representative exoplanetary temperatures (T=500K-2500K, P$\lesssim$1 atm) and H$_2$ rich atmospheres we find the median difference in cross sections introduced by various aspects of pressure broadening ($\delta$) to be $\lesssim$1\%. For higher resolutions (R$\lesssim$5000), including those attainable with JWST, we find that $\delta$ can be up to 40\%. On the other hand for very high resolution spectra (R$\sim$10$^5$) pressure broadening can introduce $\delta$$\gtrsim$100\%, reaching $\gtrsim$ 1000\% for low temperatures, (T$\lesssim$500K), high resolutions (R$\sim$10$^5$)  and high pressures (P$\sim$0.1-1 atm). Such a case could be found with instruments such as the VLT and E-ELT if cool H$_2$-dominated targets were observed. For hotter targets of T=2000K this reduces to 15\%, though this is a median over wavelength and can be found to be higher ($\gtrsim$100\%) for certain features. For spectral resolutions of R=5000, (similar to that achievable with JWST) this reduces to $\delta$=5\% for hot targets of T=2000K at P=0.1 atm. 

From this we can see that even with very high resolutions current hot Jupiter targets with temperatures of 800K-2500K will not be greatly effected by differences induced from pressure broadening with pressures of $\lesssim$0.1 atm. A more significant change is found at pressures of 1 atm or above, though current observations of exoplanetary atmospheres typically probe pressures above 1 atm \citep{madhu2012}. Data on cool targets at high resolution is currently a distant future prospect and we are unlikely to be affected by this level of uncertainty in the near future with such targets. However, even with the lower temperature end of hot Jupiter targets (T$\sim$1000K) and modern day instruments such as the VLT pressure broadening can cause discrepancies in the cross section of 30-200\% for H$_2$ dominant atmospheres with P=0.1-1 atm.

Molecular cross sections are degenerate with abundance in atmospheric models and any error in cross sections results in an uncertainty in our abundance measurements. From this work we find that for cool targets (T$\sim$500K) at high resolutions (R$\gtrsim$10$^5$) we would expect uncertainties in the abundance measurements of at least 100\% purely from the cross section inputs to atmospheric models over those that do not include pressure broadening for H$_2$ dominated atmospheres. A true spectrum involves many cross sections from an atmosphere with many layers of temperature and pressure and an observation through many optical depths which will compound this difference. For a true estimation of the difference pressure broadening will make to abundances full, rigorous atmospheric models are needed.

Cross sections have been created from a variety of sources which span different levels of completeness, i.e. the number of transitions for which line data are available, and temperature validity, i.e. the temperature up to which  the intensity values and completeness can be trusted. In this work we focus on H$_2$O as it is both currently detectable in hot Jupiters and well documented in line list sources. We find that our metric of finding the median percentage difference gives good results even when line lists have low completeness, such as the PS line list for H$_2$O. Due to this we are able to make comparisons between cross sections generated from line lists of different sources. 

Currently significant efforts are being made into obtaining data on pressure broadening parameters for molecules in different gasses. However, there are some cases where no pressure broadening data is available for certain molecules. In other cases it may be that a smaller, low temperature line list source such as HITRAN has broadening parameters for fewer transitions but a more accurate and more complete line list from sources such as ExoMol does not. To investigate what the best approach is in such situations a mean approach has been tested where broadening parameters are averaged and the mean parameter is applied to each profile. We find that, when taken across a wide wavelength range, the differences between cross sections generated in a detailed manner and those with a mean broadening parameter applied is up to 20\%, even at very high resolution (R$>$10$^5$). However when looking in detail in a narrow wavelength band individual lines may be inaccurate at higher pressure due to slight differences in the broadening parameter from the mean. Despite this, using our metric of finding the median percentage difference we find that mean broadening parameters are still useful to ascertain the magnitude of the change to the cross section pressure broadening can induce.

We also investigated the influence of different broadening agents (self, air and H$_2$) on the cross sections using H$_2$O as a case study. Generally, self broadening is significantly stronger than H$_2$ broadening, by about a factor of 4 on average across our range of pressure, temperature, and resolution. For H$_2$ broadening, which is the dominant component for giant exoplanet atmospheres that are most amenable to spectroscopy, we see a smaller effect on cross sections than that due to self broadening or air broadening. We find in our current investigations using H$_2$O that  where only self and air broadening are  available, air broadening produces a closer result to H$_2$ broadening. As with other parameters discussed above, the differences induced to cross sections due to the different broadening agents are $<$1\% for low resolution (R $\lesssim$ 100). For medium resolutions of R=5000 we find that $\delta\sim$10\% at T=500K and P=0.1 for H$_2$O in an H$_2$O atmosphere (i.e. self broadening) where as $\delta\sim$1\% for an H$_2$ atmosphere (i.e. H$_2$ broadening) . For hotter targets with higher resolution (T=1000K, R=100,000) we find that $\delta \sim$100\% for P=0.1 atm. We find that when looking at the transitions of water in a water dominated atmosphere we expect the effect of pressure broadening to be more pronounced implying that pressure broadening is likely to be very important for hotter water-rich targets. 

A partial pressure of one has been used here, assuming that only one broadening agent is present, though a  combination of agents would be more physical, particularly with He included. Further investigation could be undertaken to find at what concentration other agents affect the broadening profile shape.

A final investigation has been undertaken to assess the difference including pressure broadening in cross sections makes to the transmitted intensity through a uniform column of gas, as a function of the pressure, temperature, spectral resolution, and optical depth. Our investigation focused on an idealized column of H$_2$-dominated gas with H$_2$O as the only absorber in the HST WFC3 G141 bandpass (1.1-1.7 $\mu$m). The results follow the general trends of how each of these parameters influence the cross sections themselves, as discussed above. For low resolutions (R$\lesssim$100), we find the relative change in transmittance ($\Delta I_{\lambda}$) to be $\lesssim$1\% across the HST WFC3 G141 bandpass (1.1-1.7 $\mu$m) for almost the entire range of parameters of relevance to exoplanetary atmospheres. For representative parameters of $\tau<1$, P$<$0.1 atm, and T$>$1000K, $\Delta I_{\lambda}$ can be up to 6\% for a JWST-like medium resolution of R = 5000, and 75\% for a VLT-like very high resolution of R = $10^5$. $\Delta I_{\lambda}$ can be even higher for higher T, lower T, and larger $\tau$. While for R $\lesssim$ 5000 the $\Delta I_{\lambda}$ are still below $\sim$100\%, for very high resolutions (R$\sim$10$^5$) $\Delta I_{\lambda}$ can be as high as $\sim$2000\%.

Ultimately, our present work suggests that incorporating pressure broadening to compute molecular cross sections for atmospheric models will be necessary depending on the desired accuracy in molecular abundance estimates retrieved from the spectra. Across all the various factors considered in this work, for low resolution observations (R$\lesssim$100) of exoplanetary spectra that are currently possible, e.g. with HST, the median differences in cross section induced due to accurate pressure broadening is found to be $\lesssim $1\%. For medium resolutions (R$\sim$5000), similar to those possible with JWST, the differences are expected to be at the $\lesssim$40\% level. For very high resolution spectra (R$\sim$$10^5$), that are possible with current and future large ground-based telescopes such as VLT and E-ELT, significantly higher differences are possible of 100\% or  even much larger, depending on other factors discussed above.

With atmospheric characterisation becoming an ever more important part of exoplanet research we can begin to see that pressure broadening will impact us in the future. With medium and high resolution spectra of exoplanets, both hot and cool, we can expect our abundance measurements to be affected in some way. Beyond that we may be able to detect and characterise the pressure in the atmospheres of other planets by finding regions of wavelength space particularly affected and using high resolution spectra. This goal would be difficult to achieve even with a wealth of molecular data at our disposal as signal to noise ratios for such data are likely to be low. The structure and dynamics of a full atmosphere leads  to a convolution of many profiles making characterisation of pressure broadening exceedingly difficult. Other factors such as wind speed and Doppler broadening provide further barriers. These will be the ultimate challenges of the future when we will eventually be able to conduct very high resolution spectroscopy of cool low mass exoplanets.


\section*{Acknowledgements}
CH acknowledges the support from the Science and Technology Facilities Council (STFC) for PhD funding. We thank the groups that have spent time developing molecular line lists, particularly ExoMol, HITEMP and HITRAN, without which this work would not be possible. We thank Richard Freedman, Iouli Gordon, Jonathan Tennyson, and Sergey Yurchenko for helpful discussions. We would like to thank the anonymous reviewer for a very helpful review.

\bibliographystyle{mn2e}
\bibliography{draft}

\section{Appendix}
\subsection{Available Line List Sources}
\label{LineListSources}
Available sources for line lists are listed in Table~\ref{sources} with each molecule that is offered. Some lists such as those from HITRAN, HITEMP and GEISA offer pressure broadening parameters as well as intensity and wavelength for transitions. Here we give the number of transitions for each list for each molecule. Where two sources are available it is on the whole better to use one with more transitions as it is more complete, though there is also temperature validity to consider. We have used HITEMP's water line list as the most accurate in this work as it is an update of the BT2 line list. ExoMol, however, does offer the most complete line lists with the highest temperature validity range so where possible we recommend using their line lists though they do not currently provide pressure broadening parameters.

\begin{center}
  \begin{table*}
    \centering
    \begin{tabular}{|c|c|c|c|c|c|c|c|}
      \textbf{Molecule} & \textbf{HITRAN} & \textbf{HITEMP} & \textbf{GEISA} & \textbf{ExoMol} & \textbf{Yueqi} & \textbf{Other} & \emph{(Source)}\\\hline

      AlO and Isotopes	&	-	&	-	&	-	&	5000000	&	-	&	-	&	-	\\
      BeH	&	-	&	-	&	-	&	7858	&	-	&	-	&	-	\\
      C2	&	-	&	-	&	-	&	-	&	47570	&	-	&	-	\\
      C2H2	&	12613	&	-	&	11340	&	-	&	-	&	-	&	-	\\
      C2H4	&	18097	&	-	&	18378	&	-	&	-	&	-	&	-	\\
      C2H6	&	43592	&	-	&	28439	&	-	&	-	&	-	&	-	\\
      C2HD	&	-	&	-	&	15512	&	-	&	-	&	-	&	-	\\
      C2N2	&	-	&	-	&	2577	&	-	&	-	&	-	&	-	\\
      C3H4	&	-	&	-	&	19001	&	-	&	-	&	-	&	-	\\
      C3H8	&	-	&	-	&	8983	&	-	&	-	&	-	&	-	\\
      C4H2	&	124126	&	-	&	119480	&	-	&	-	&	-	&	-	\\
      C6H6	&	-	&	-	&	9797	&	-	&	-	&	-	&	-	\\
      CaH	&	-	&	-	&	-	&	26980	&	6000	&	-	&	-	\\
      CF4	&	60033	&	-	&	291	&	-	&	-	&	-	&	-	\\
      CH and Isotopes	&	-	&	-	&	-	&	-	&	53086	&	-	&	-	\\
      CH3Br	&	18692	&	-	&	36911	&	-	&	-	&	-	&	-	\\
      CH3Cl	&	107642	&	-	&	18344	&	-	&	-	&	-	&	-	\\
      CH3CN	&	-	&	-	&	200	&	-	&	-	&	-	&	-	\\
      CH3CN	&	3572	&	-	&	-	&	-	&	-	&	-	&	-	\\
      CH3D	&	-	&	-	&	49237	&	-	&	-	&	-	&	-	\\
      CH3OH	&	-	&	-	&	19897	&	-	&	-	&	-	&	-	\\
      CH3OH	&	19897	&	-	&	-	&	-	&	-	&	-	&	-	\\
      CH4	&	336830	&	-	&	240858	&	10 $\times 10^{10}$	&	-	&	-	&	-	\\
      ClO	&	5721	&	-	&	7230	&	-	&	-	&	-	&	-	\\
      ClONO2	&	21988	&	-	&	356899	&	-	&	-	&	-	&	-	\\
      CN	&	-	&	-	&	-	&	-	&	195120	&	-	&	-	\\
      CO	&	1019	&	113631	&	13515	&	-	&	-	&	-	&	-	\\
      CO2	&	169292	&	11193608	&	413524	&	-	&	-	&	573881316	&	\cite{CDSD-4000}	\\
      COF2	&	168793	&	-	&	70904	&	-	&	-	&	-	&	-	\\
      CP	&	-	&	-	&	-	&	-	&	-	&	-	&	-	\\
      CrH	&	-	&	-	&	-	&	-	&	13825	&	-	&	-	\\
      CS	&	5129	&	-	&	-	&	-	&	-	&	-	&	-	\\
      FeH	&	-	&	-	&	-	&	-	&	93040	&	-	&	-	\\
      FeH	&	-	&	-	&	-	&	-	&	-	&	204688	&	\cite{Bernath}	\\
      GeH4	&	-	&	-	&	824	&	-	&	-	&	-	&	-	\\
      H2	&	4017	&	-	&	-	&	-	&	-	&	-	&	-	\\
      H2CO	&	-	&	-	&	37050	&	-	&	-	&	-	&	-	\\
      H2O	&	142045	&	114241164	&	67504	&	505000000	&	-	&296000000	&	\cite{PS}	\\
      H2O2	&	126983	&	-	&	126983	&	-	&	-	&	-	&	-	\\
      H2S	&	36561	&	-	&	20788	&	-	&	-	&	-	&	-	\\
      HBr	&	3039	&	-	&	1294	&	-	&	-	&	-	&	-	\\
      HC3N	&	180332	&	-	&	179347	&	-	&	-	&	-	&	-	\\
      HCl	&	11879	&	-	&	533	&	-	&	2588	&	-	&	-	\\
      HCN	&	2955	&	-	&	81889	&	34418408	&	-	&	-	&	-	\\
      HCOOH	&	62684	&	-	&	62684	&	-	&	-	&	-	&	-	\\
      HDO	&	-	&	-	&	-	&	700000000	&	-	&	-	&	-	\\
      HF	&	10073	&	-	&	107	&	-	&	-	&	-	&	-	\\
      HI	&	3161	&	-	&	806	&	-	&	-	&	-	&	-	\\
      HNC	&	-	&	-	&	5619	&	-	&	-	&	-	&	-	\\
      HNO3	&	-	&	-	&	669988	&	-	&	-	&	-	&	-	\\
      HNO3	&	903854	&	-	&	-	&	-	&	-	&	-	&	-	\\
      HO2	&	38804	&	-	&	38804	&	-	&	-	&	-	&	-	\\
      HOBr	&	2177	&	-	&	-	&	-	&	-	&	-	&	-	\\
      HOCl	&	8877	&	-	&	17862	&	-	&	-	&	-	&	-	\\
      Kcl and Isotopes	&	-	&	-	&	-	&	7000000	&	-	&	-	&	-	\\
      LiH	&	-	&	-	&	-	&	18982	&	-	&	-	&	-	\\
      MgH and Isotopes	&	-	&	-	&	-	&	6716	&	-	&	-	&	-	\\
      N2	&	1107	&	-	&	120	&	-	&	-	&	-	&	-	\\
      N2O	&	33074	&	-	&	50633	&	-	&	-	&	-	&	-	\\
      NaCl and Isotopes	&	-	&	-	&	-	&	5000000	&	-	&	-	&	-	\\
      NaD	&	-	&	-	&	-	&	167224	&	-	&	-	&	-	\\
      NaH	&	-	&	-	&	-	&	79898	&	-	&	-	&	-	\\
      NH	&	-	&	-	&	-	&	-	&	10425	&	-	&	-	\\
      NH3	&	45302	&	-	&	29082	&	1138 323 351	&	-	&	-	&	-	\\
      NO	&	103710	&	115610	&	105079	&	-	&	-	&	-	&	-	\\
      \hline

      \label{sources}

    \end{tabular}
  \end{table*}
\end{center}

\begin{center}
  \begin{table*}
    \centering
    \begin{tabular}{|c|c|c|c|c|c|c|c|}
      \textbf{Molecule} & \textbf{HITRAN} & \textbf{HITEMP} & \textbf{GEISA} & \textbf{ExoMol} & \textbf{Yueqi} & \textbf{Other} & \emph{(Source)}\\\hline
      NO+	&	1206	&	-	&	1206	&	-	&	-	&	-	&	-	\\
      NO2	&	104223	&	-	&	104223	&	-	&	-	&	-	&	-	\\
      O	&	2	&	-	&	-	&	-	&	-	&	-	&	-	\\
      O2	&	1787	&	-	&	6428	&	-	&	-	&	-	&	-	\\
      O3	&	261886	&	-	&	389378	&	-	&	-	&	-	&	-	\\
      OCS	&	15618	&	-	&	33809	&	-	&	-	&	-	&	-	\\
      OH	&	30772	&	41557	&	42866	&	-	&	-	&	-	&	-	\\
      PH	&	-	&	-	&	-	&	1.68$\times 10^{10}$ 	&	-	&	-	&	-	\\
      PH3	&	22189	&	-	&	20364	&	-	&	-	&	-	&	-	\\
      PN and Isotopes	&	-	&	-	&	-	&	700000	&	-	&	-	&	-	\\
      ScH	&	-	&	-	&	-	&	1152826	&	-	&	-	&	-	\\
      SF6	&	2889065	&	-	&	92398	&	-	&	-	&	-	&	-	\\
      SiO and Isotopes	&	-	&	-	&	-	&	1784965	&	-	&	-	&	-	\\
      SO2	&	72460	&	-	&	68728	&	-	&	-	&	-	&	-	\\
      SO3	&	10188	&	-	&	-	&	174674257	&	-	&	-	&	-	\\
      TiH	&	-	&	-	&	-	&	-	&	181080	&	157430	&	\cite{Bernath}	\\
      TiO	&	-	&	-	&	-	&	-	&	-	&	8325354	&	\cite{kurucz}	\\
      \hline

    \end{tabular}
    \caption{A list of all available line list sources for different molecules with the number of lines available in each. In general the more complete line lists are preferable not only as more features are evaluated within the molecular spectrum but because the continuum of low intensity lines is better approximated.}
    \label{sources}

  \end{table*}
\end{center}

\label{lastpage}
\nocite{*}

\end{document}